# *Quantum Geometrodynamics Revived*
# *I. Classical Constraint Algebra*

Thorsten Lang[*,1] and Susanne Schander[†,2]

[1]Institute for Quantum Gravity, FAU Erlangen – Nürnberg, Staudtstraße 7/B2, 91058 Erlangen, Germany
[2]Perimeter Institute, 31 Caroline St N, Waterloo, ON N2L 2Y5, Canada

June 23, 2023

## Abstract

In this series of papers, we present a set of methods to revive quantum geometrodynamics which encountered numerous mathematical and conceptual challenges in its original form promoted by Wheeler and De Witt. In this paper, we introduce the regularization scheme on which we base the subsequent quantization and continuum limit of the theory. Specifically, we employ the set of piecewise constant fields as the phase space of classical geometrodynamics, resulting in a theory with finitely many degrees of freedom of the spatial metric field. As this representation effectively corresponds to a lattice theory, we can utilize well–known techniques to depict the constraints and their algebra on the lattice. We are able to compute the lattice corrections to the constraint algebra. This model can now be quantized using the usual methods of finite–dimensional quantum mechanics, as we demonstrate in the following paper. The application of the continuum limit is the subject of a future publication.

## Contents

[*]thorsten.lang@fau.de
[†]sschander@perimeterinstitute.ca

## 1   Introduction

The quest for a viable theory of quantum gravity is a longstanding and yet still controversially discussed problem of theoretical physics. Numerous promising avenues have emerged over the years, including string theory (Polchinski 2007), loop quantum gravity (LQG) (Ashtekar and Pullin 2017), asymptotic safety (Reuter and Saueressig 2019), and causal dynamical triangulations (CDT) (Loll 2020), among others. However, despite the existence of these varied approaches, significant questions and challenges persist in attaining a widely embraced solution within the scientific community.

One of the pionieering endeavors in the quantization of gravity was the Hamiltonian quantum geometrodynamical approach which employs a (3 + 1) split of spacetime in which the spatial metric fields $q_{ab}(x)$ and their conjugate momenta $p^{cd}(x)$ represent the phase space of the theory. A comprehensive review of this approach can be found in (Arnowitt et al. 2008). Upon analyzing the dynamics of the theory, it becomes apparent that the system is entirely constrained, reflecting its invariance under general coordinate transformations. Following Dirac's approach to constrained systems (Dirac 1958), the resulting set of four constraints – the Hamiltonian and the three diffeomorphism constraints – must be preserved under the flow generated by the constraints themselves to ensure the consistency of the theory. At the classical level, this consistency condition is satisfied as the Poisson algebra of the gravitational constraints vanishes weakly on the constraint hypersurface.

The quantization of this theory presents a formidable challenge as was already noted by DeWitt (1967) (see also references therein). To quantize constrained systems, one would typically follow the procedure outlined by Dirac (1964) wherein the constraints are imposed as quantum operators on physical states of an appropriate Hilbert space. Regrettably, the non–polynomial structure and the presence of interaction terms within the Hamiltonian constraint of gravity complicate matters, rendering the construction of a well–defined kinematical Hilbert space for the theory elusive. Consequently, representing quantum constraints and establishing an inner product for physical states on such a hypothetical Hilbert space remain ambiguous tasks. Additional and interrelated concerns are the possibility of anomalies in the quantum constraint algebra, which could undermine the theory's consistency, the factor ordering problem and the definition of a suitable regularization scheme (Tsamis and Woodard 1987). Other fundamental issues related to quantum geometrodynamics include the possibility of signature change of the metric tensor and the problem of time. We refer to the references by Isham (1991) and Kiefer (2007) for in–depth discussions of the above–mentioned problems.

To address these challenges, it appears reasonable to explore and adapt successful tools utilized in other field–theoretic frameworks, such as the electroweak or strong forces. In these contexts, a common strategy involves introducing a regularization scheme to tackle ultraviolet divergences and subsequently seeking the continuum limit through renormalization procedures.

The idea of applying regularization (and potentially renormalization) to gravity is not new, and various popular regularization schemes have emerged. Some of them utilize simplicial building blocks to truncate spacetime. Examples include Regge calculus (Regge 1961), CDT and spinfoams (Rovelli and Vidotto 2020). In the asymptotic safety approach, a common practice is to employ a UV cutoff by restricting the momentum scale.

In order to gain insight into the corresponding quantum theories, both in the lattice approaches to the standard model and in gravity, most of the aforementioned methods rely on a path integral formalism. This choice appears natural when the objective is to compute transition amplitudes between different states in the theory. However, the path integral approach to quantum gravity encounters several difficulties, such as determining a suitable measure for the path integral and addressing the question of appropriate quantum states for computing transition amplitudes. The definition of a physical Hilbert space typically necessitates a Hamiltonian treatment of the theory.

In the context of lattice quantum chromodynamics, Kogut and Susskind (1975) introduced a Hamiltonian formulation that utilizes spatial lattice regularization. Remarkably, Creutz (1985) demonstrated the equivalence of this approach with the path integral formulation commonly employed in lattice gauge theories.

Regarding gravity, various regularization approaches within the Hamiltonian framework have been explored, as reviewed by Loll (1998a) and Williams (2009). Among these, one noteworthy approach proposed by Gambini and Pullin (2003) employs Ashtekar variables and simplicial building blocks. However, to the best of our knowledge, nobody has ever investigated a spatial regularization scheme and subsequent continuum limit of quantum geometrodynamics in its Hamiltonian formulation.

In our series of papers, of which this is the first one, we lay the groundwork for a new proposal for Hamiltonian quantum gravity following the above mentioned principles. We proceed along the lines of standard lattice gauge theories in that we first regularize the theory and then aim at performing a well–defined continuum limit. Following the Hamiltonian approach, we perform a $3 + 1$ split (or more generically, an $n + 1$ split) such that our basic variables are the spatial metric components. Instead of introducing connection or loop variables as is done in LQG, we stay close to the original geometrodynamics approach and treat the metric field as fundamental. In this paper, we start by defining our approach to this lattice theory of gravity, and by evaluating the corresponding constraint algebra on the lattice. All considerations here are purely classical.

More precisely, our approach to gravity on the lattice starts by defining the phase space as the set of canonical fields that are piecewise constant with respect to a regular lattice embedded in the spatial manifold $\sigma$. This amounts to assigning a finite number of degrees of freedom to each hypercube of the lattice, specifically, we have $n(n + 1)/2$ parameters associated with the symmetric metric tensor for each hypercube or at each lattice site. We will also explain how this restriction to piecewise constant fields can gradually be lifted. For simplicity, we assume the lattice to admit periodic boundary conditions. We replace integrals by weighted sums and partial derivatives by finite (central) differences. We choose a representation of the constraints on the lattice that resembles the continuum case closely and which simplifies computations. We are then able to recover the hypersurface deformation algebra on the lattice augmented by additional terms proportional to the lattice spacing $\epsilon$. These additional terms are regular in the sense that they depend on regular functions of the phase space variables on their domain of definition. These anomalies also include couplings between different lattice sites, i.e., the algebra is non–local. We emphasize that the appearance of anomalies is expected since the approach breaks diffeomorphism invariance by defining the theory on the lattice. While such anomalies appear certainly problematic in fundamentally discrete approaches to quantum gravity, we believe that such anomalies are just a logical consequence of the regularization procedure and should disappear after taking a suitable continuum limit. We note that computations of the commutator algebra have also been undertaken within other frameworks to lattice (quantum) gravity (Bander 1987; Bonzom and Dittrich 2013; Friedman and Jack 1986; Loll 1998b; Piran and Williams 1986) (see also references therein).

In the second paper of this series (Lang and Schander 2023a), we are able to construct a rigorously defined Hilbert space at each individual lattice site. More significantly, the total Hilbert space can be represented as a tensor product of the Hilbert spaces at each lattice site, thereby ensuring a consistent and well–defined framework for the quantum theory under consideration. Furthermore, we are able to establish the positive definiteness of the metric tensor in our quantum representation. For a brief and condensed summary of our current findings, we refer to the first paper of the series (Lang and Schander 2023b).

Finally, let us mention that our approach can be viewed as a quantization of numerical relativity since we are using similar methods to define a lattice version of general relativity (Alcubierre 2008).

With this, let us come to the structure of this paper: In section 2, we briefly review the Hamiltonian approach to classical gravity in the continuum and introduce its constraint algebra. Section 3 provides the basics of our approach to considering gravity on the lattice and introduces the lattice calculus that we will be using. In section 4, we provide some useful formulas of Riemannian "geometry" on the lattice. Section 5 specifies our representation of the lattice constraints and computes their algebra. Finally, section 6 closes with a summary and a discussion of our findings.

*Notation*

We set $c = 16\pi G = 1$. Lower case latin letters $a, b, c, ...$ refer to spatial indices ranging from 1 to $n$ while lower case greek letters $\mu, \nu, \rho, ...$ denote spacetime indices and hence run from 0 to $n$.

## 2   Constraints in the Continuum

Following Arnowitt et al. (1959), the canonical theory of gravity in the continuum rests on introducing a spatial metric $q_{ab}$ on the $n$–dimensional spatial hypersurfaces $\sigma$ of the $(n+1)$–dimensional spacetime manifold $\mathcal{M} = \mathbb{R} \times \sigma$. Together with the lapse function $N$ and the shift vector field $\mathcal{N}^b$, the $(n + 1)$–dimensional Lorentzian metric $g_{\mu\nu}$ is then given by

$$g_{00} = -N^2 + \mathcal{N}_b \mathcal{N}^b, \qquad g_{0b} = g_{b0} = \mathcal{N}_b, \qquad g_{bc} = q_{bc}. \tag{2.1}$$

It is useful to introduce the second fundamental form $K_{bc} := (\dot{q}_{bc} - 2D_{[b}\mathcal{N}_{c]})/(2N)$ where the dot denotes a time derivative and $D_b$ is the covariant derivative on the spatial hypersurfaces. In terms of these quantities, together with the determinant of the spatial metric $q := \det q$, the Lagrangian of general relativity becomes (Wald 1984)

$$L = \int_\sigma d^n x \, N\sqrt{q}\left(K_{bc}K^{bc} - K^2 + R\right) \tag{2.2}$$

where $K = K_{bc}q^{bc}$, $R$ is the curvature scalar of the spatial hypersurfaces and we set the cosmological constant $\Lambda = 0$. Note that we discard two dynamically irrelevant boundary terms to derive equation (2.2) (DeWitt 1967). The momenta conjugate to $N$, $\mathcal{N}_b$ and $q_{bc}$ are respectively given by

$$\pi := \frac{\partial \mathcal{L}}{\partial \dot{N}} = 0, \quad \pi_b := \frac{\partial \mathcal{L}}{\partial \dot{\mathcal{N}}^b} = 0, \quad p^{bc} := \frac{\partial \mathcal{L}}{\partial \dot{q}_{bc}} = \sqrt{q}\left(K^{bc} - Kq^{bc}\right) \tag{2.3}$$

where $\mathcal{L}$ is the Lagrange density defined by $L = \int d^n x \, \mathcal{L}$. The first two equations represent primary constraints according to Dirac (1964) and express the fact that the velocities $\dot{N}$ and $\dot{\mathcal{N}}^b$ can be chosen arbitrarily. A Legendre transformation using the above momenta yields the total Hamiltonian

$$\mathbf{H} = \int_\sigma d^n x \left(\lambda \pi + \lambda^b \pi_b + N\mathcal{H} + \mathcal{N}^b \mathcal{C}_b\right) \tag{2.4}$$

where $\lambda$ and $\lambda^b$ are Lagrange multipliers (Dirac 1964), and the Hamilton constraint density $\mathcal{H}$ and the diffeomorphism constraint densities $\mathcal{C}_b$ are given by (Thiemann 2007b)

$$\mathcal{H} := \frac{1}{\sqrt{q}}\left(q_{bd}q_{ce} - \frac{1}{n-1}q_{bc}q_{de}\right)p^{bc}p^{de} - \sqrt{q}R, \tag{2.5}$$

$$\mathcal{C}_b := -2D_d p^d{}_b = -2(q_{bc}p^{cd})_{,d} + q_{cd,b}p^{cd} \tag{2.6}$$

where we used that $p^{cd}$ is a tensor density of weight one in order to express its covariant derivative in terms of a partial derivative. Defining the smeared versions of the Hamilton and the diffeomorphism constraints

$$H(N) := \int_\sigma d^n x \, N\mathcal{H}, \quad C(\mathcal{N}) := \int_\sigma d^n x \, \mathcal{N}^b \mathcal{C}_b, \tag{2.7}$$

the Dirac algebra (Dirac 1964) which encodes the non–trivial dynamics of the system is given by

$$\{C(\mathcal{N}_1), C(\mathcal{N}_2)\} = -C(\mathcal{L}_{\mathcal{N}_1}\mathcal{N}_2), \tag{2.8}$$

$$\{C(\mathcal{N}), H(N)\} = -H(\mathcal{L}_{\mathcal{N}}N), \tag{2.9}$$

$$\{H(N_1), H(N_2)\} = -C(V(q, N_1, N_2)) \tag{2.10}$$

where $\mathcal{L}_g f$ is the Lie derivative of the (scalar, vector or tensor) function $f$ along the vector field $g$, and $V(q, N_1, N_2)$ is the vector field defined by $V^b := q^{bc}(N_1 \partial_c N_2 - N_2 \partial_c N_1)$.

## 3 Lattice Theory

Our aim is to formulate a theory of gravity with a finite number of spatial degrees of freedom. To accomplish this, we limit the theory's phase space to the set of piecewise constant functions (and tensors) as defined below. This allows us to use the tools from conventional lattice theories while formally physical fields remain defined on the entire spatial manifold. The next step in our approach involves gradually lifting this restriction.

Let us illustrate this procedure in one spatial dimension. Let $[a,b) \subset \mathbb{R}$ be an interval on the real line. Let us assume that $b - a > 0$ is an $N$'th multiple of $\epsilon \in \mathbb{R}^+$, in particular $b - a = N\epsilon$. We divide $[a,b)$ into $N$ intervals all of the same length $\epsilon$, and label each of these intervals $I_X$ by an entire number $X \in 0, \ldots, N-1$. More precisely, we define

$$I_X := [a + X\epsilon, a + (X+1)\epsilon) \subset \mathbb{R}. \tag{3.1}$$

The idea is to confine the space of admissible functions $f : [a,b) \mapsto \mathbb{R}$ to those that are constant on each $I_X$ with function value $f^X$ on $I_X$, i.e.,

$$f(x) := \sum_{X=0}^{N-1} f^X \begin{cases} 1, & \text{if } x \in I_X \\ 0, & \text{otherwise} \end{cases} = \sum_{X=0}^{N-1} f^X \chi_X(x), \tag{3.2}$$

where we introduced the characteristic function $\chi_X(x)$ which is one if $x \in I_X$ and zero otherwise. The integral of $f(x)$ is given by

$$\int_a^b \mathrm{d}x\, f(x) = \sum_{X=0}^{N-1} f^X \int_a^b \mathrm{d}x\, \chi_X(x) = \epsilon \sum_{X=0}^{N-1} f^X, \tag{3.3}$$

where by integrating out the characteristic function, we obtain an additional factor $\epsilon$ in the final step. The integral reduces to a sum, very similar to how one defines "integration" for lattice theories. In particular, one can think of each interval $I_X$ as hosting one lattice point $X$ and $\epsilon$ is the lattice constant. This view point allows us to consider our reduced theory as a lattice theory and simplifies the mathematical description. We emphasize however that $f(x)$ is still defined on the entire interval $[a,b)$.

In this work, we focus on the case of periodic boundary conditions which enables us to study the entire spacetime within a finite region. Specifically, in the one dimensional case, we make the identification

$$f^X = f^{Y \bmod N} \qquad \forall Y \in \mathbb{Z}, X \in 0, \ldots, N-1, \tag{3.4}$$

for any piecewise constant function $f^X$ on $[a,b)$. In particular, we have that $f^N = f^0$ and $f^{N+1} = f^1$.

In $n$ (spatial) dimensions one can play a similar game. While in one dimension the lattice points sit on the real line, we define a hypercubical lattice for $n$ dimensions in close analogy by

$$\Lambda := \{(X_1, \ldots, X_n) : X_i \in 0, \ldots, N-1\ \forall i = 1, \ldots, n\}. \tag{3.5}$$

The intervals become hypercubes $I_{X_1 \ldots X_n}$ and a piecewise constant function is constant over each of the hypercubes respectively. Hence, a function $f$ reduces to

$$f(x) := \sum_{X_1 \ldots X_n = 0}^{N-1} f^{X_1 \ldots X_n} \chi_{X_1 \ldots X_n}(x), \tag{3.6}$$

while its integral over $[a,b)^n$ is given by

$$\int_{[a,b)^n} \mathrm{d}^n x\, f(x) = \epsilon^n \sum_{X_1\ldots X_n=0}^{N-1} f^{X_1\ldots X_n}. \tag{3.7}$$

Again, the value $f^{X_1\ldots X_n}$ can be thought of being assigned to the lattice point $(X_1, \ldots, X_n)$. This concept leads to the idea of a lattice function. As for the one dimensional case, we specialize to periodic boundary conditions.

**Definition 3.1** (Lattice Function). A real–valued function $f$ on the lattice $\Lambda$ with periodic boundary conditions is given by the assignment

$$f : \Lambda \to \mathbb{R} : (X_1, \ldots, X_n) \mapsto f^{X_1\ldots X_n}, \tag{3.8}$$

where we require that

$$f^{X_1\ldots X_n} = f^{(Y_1 \bmod N)\ldots(Y_n \bmod N)} \quad \forall Y_i \in \mathbb{Z},\ X_i \in 0,\ldots,N-1\ \forall i \in 1,\ldots,n. \tag{3.9}$$

The space of such functions will be denoted by $\mathcal{F}(\Lambda, \mathbb{R})$.

We also seek to define finite differences of lattice functions in order to replace the continuum derivatives. To accomplish this, we employ a "central" finite difference.

**Definition 3.2** (Central Difference). The central difference operator $\Delta_a$ acts on a lattice function $f \in \mathcal{F}(\Lambda, \mathbb{R})$ with periodic boundary conditions according to

$$(\Delta_a f)^{X_1\ldots X_n} := \frac{f^{X_1\ldots(X_a+1 \bmod N)\ldots X_n} - f^{X_1\ldots(X_a-1 \bmod N)\ldots X_n}}{2\epsilon} \tag{3.10}$$

with $X_i \in 0,\ldots,N-1$ and for all $i \in 1,\ldots,n$.

The following definition also proves to be useful.

**Definition 3.3** (Lattice Shift Operators). The shift operator $T_a$ and its inverse $T_a^{-1}$ act on a lattice function $f \in \mathcal{F}(\Lambda, \mathbb{R})$ by shifting its value according to

$$(T_a f)^{X_1\ldots X_n} := f^{X_1\ldots X_a+1 \bmod N\ldots X_n}, \tag{3.11}$$

$$\left(T_a^{-1} f\right)^{X_1\ldots X_n} := f^{X_1\ldots X_a-1 \bmod N\ldots X_n}. \tag{3.12}$$

Omitting lattice indices, this allows us to write the central difference operator as

$$\Delta_a f = \frac{T_a f - T_a^{-1} f}{2\epsilon}. \tag{3.13}$$

On the lattice, the product rule of differentiation used in continuous calculus must be augmented by a lattice term proportional to the lattice constant.

**Definition 3.4** (Lattice Product Rule). The difference operator $\Delta$ acts on the product of two lattice functions $f, g \in \mathcal{F}(\Lambda, \mathbb{R})$ according to

$$\Delta_a(f \cdot g) = (\Delta_a f) \cdot g + f \cdot (\Delta_a g) + \epsilon\, \mathcal{O}_a(f, g), \tag{3.14}$$

where the additional lattice contribution can be written using the so–called forward and backward lattice differences $\Delta^{\mathrm{f},\epsilon}$ and $\Delta^{\mathrm{b},\epsilon}$ as

$$\mathcal{O}_a(f, g) := \frac{1}{2}\big((\Delta_a^{\mathrm{f}} f)(\Delta_a^{\mathrm{f}} g) - (\Delta_a^{\mathrm{b}} f)(\Delta_a^{\mathrm{b}} g)\big) \tag{3.15}$$

where

$$(\Delta_a^{\mathrm{b}} f) := \frac{T_a f - f}{\epsilon}, \quad (\Delta_a^{\mathrm{f}} f) := \frac{f - T_a^{-1} f}{\epsilon}. \tag{3.16}$$

The central, forward and backward differences on the lattice are linked by the assignment

$$\Delta_a f = \frac{\Delta_a^{\mathrm{f}} f + \Delta_a^{\mathrm{b}} f}{2}. \tag{3.17}$$

Whether one chooses the central, forward or backward difference is a question of personal taste. In the continuum limit, all three choices should reduce to the same usual derivative operator. Likewise, when sending $\epsilon \to 0$, the additional lattice contributions from the product rule should disappear. In this work, we use the central difference as it simplifies our computations, in particular, it translates the continuous integration by parts to the following summation rule.

**Lemma 3.5** (Summation by Parts). *Let $f, g \in \mathcal{F}(\Lambda, \mathbb{R})$ be two periodic lattice functions. Then, the following rule for shifting the central difference operator holds:*

$$\sum (\Delta_a f) g = - \sum f (\Delta_a g), \tag{3.18}$$

*where the sum runs from 0 to $N-1$ for all $X_1, \ldots, X_n$ and $f$ and $g$ depend on these lattice coordinates.*

The proof is straightforward, and we will not detail it here. For our goal of defining general relativity on the lattice, we need a notion of tensors of different types and of lattice tensor "fields". While the conventional definitions of such objects employing tangent and cotangent spaces on a smooth manifold obviously fail on the lattice, it is still possible to define a tensor at each lattice point.

**Definition 3.6** (Lattice Tensor). A tensor $T$ of type $(r, s)$ at the lattice point $(X_1 \ldots X_n)$ is a multi-linear map

$$T^{X_1 \ldots X_n} : V^* \times \cdots \times V^* \times V \times \cdots \times V \to \mathbb{R} \tag{3.19}$$

which maps $r$ dual vectors $v^1, \ldots, v^r \in V^*$ of an $n$–dimensional (dual) real vector space $V^*$ and $s$ vectors $v_1, \ldots, v_s \in V$ of an $n$–dimensional real vector space $V$ onto $\mathbb{R}$. The space of $(r, s)$ tensors is denoted by $\mathcal{T}_{(r,s)}$.

As such, the usual index notation $(T^{X_1...X_n})^{a_1...a_r}{}_{b_1...b_s}$ can be used. The conventional contraction of indices, the Einstein summation rule and the lowering (raising) of indices with the (inverse) metric tensor still apply pointwise. A "lattice tensor field" is a collection of tensor fields at the lattice sites.

**Definition 3.7** (Lattice Tensor Field). A lattice tensor field of type $(r, s)$ on the periodic lattice $\Lambda$ is a map

$$T : \Lambda \to \mathcal{T}_{(r,s)} : X_1 \ldots X_n \mapsto T^{X_1 \ldots X_n} \tag{3.20}$$

that assigns an $(r, s)$ tensor $T^{X_1...X_n}$ to each lattice site $X_1 \ldots X_n \in \Lambda$.

As before, we omit the lattice site indices $X_1 \ldots X_n$ whenever possible.

In the continuum limit where $\epsilon \to 0$, the usual geometric interpretation of coordinate transformations and quantities on a smooth manifold should be restored.

Similarly, we would like to have a notion of tensor densities although the usual intuition about their transformation when passing to another coordinate system is, again, completely lost on the lattice. Nevertheless, we will label the square root of the lattice metric tensor's determinant as well as the metric's conjugate momentum as "lattice tensor densities of weight one" and introduce respective rules of lattice calculus for them.

## 4 Formulas in Riemannian Lattice Geometry

To aid the reader's comprehension in the subsequent sections, we present a set of formulas in Riemannian lattice "geometry" that will be employed. The next section utilizes these formulas to compute the constraint algebra on the lattice. Therefore, the reader may choose to peruse this section at a later time if desired. To enhance the readability of this section, we moved the proofs of the lemmas to appendix A.

We start by introducing the lattice metric $q_{ab}$ and its inverse $q^{bc}$ both defined in accordance with the continuum metric but limited to the lattice sites and satisfying the usual relations $q_{ab}q^{bc} = \delta_a^c$ and $q_{ab}q^{ab} = n$. Together with the central difference (definition 3.2), this leads us to define the lattice Christoffels symbol according to

$$\Gamma^a{}_{bc} := \frac{1}{2} q^{ad} (\Delta_b q_{dc} + \Delta_c q_{bd} - \Delta_d q_{bc}), \tag{4.1}$$

and which is symmetric with respect to its lower indices. For the following, it is useful to define the notion of a "covariant difference operation" on the lattice. The covariant difference operator on the lattice can be expressed in a manner that parallels its counterpart in the continuum.

**Definition 4.1** (Lattice Covariant Difference). The covariant difference $D_a$ of a lattice tensor density $T$ of type $(r, s)$ and of weight $w$ is given by

$$D_a T^{b_1...b_r}{}_{c_1...c_s} := \Delta_a T^{b_1...b_r}{}_{c_1...c_s} + \Gamma^{b_1}{}_{da} T^{db_2...b_r}{}_{c_1...c_s} + \cdots + \Gamma^{b_r}{}_{da} T^{b_1...b_{r-1}d}{}_{c_1...c_s} \tag{4.2}$$
$$- \Gamma^d{}_{c_1 a} T^{b_1...b_r}{}_{dc_2...c_s} - \cdots - \Gamma^d{}_{c_s a} T^{b_1...b_r}{}_{c_1...c_{s-1}d} - w \Gamma^d{}_{da} T^{b_1...b_r}{}_{c_1...c_s}.$$

Note that by a simple algebraic calculation, one can show that $D_a q_{bc} = 0$ while $D_a q^{bc}$ fails to be zero on the lattice.

**Lemma 4.2** (Covariant Difference of Inverse Metric). *The covariant difference of the inverse metric vanishes up to lattice terms, i.e.,*

$$D_a q^{bc} = \epsilon \mu D q_a{}^{bc}, \quad \text{where} \quad \mu D q_a{}^{bc} := -q^{bd} \mathcal{O}_a(q_{de}, q^{ec}). \tag{4.3}$$

Similar results hold for the metric determinant and its square root as we propose in the following lemmas 4.3 and 4.4.

**Lemma 4.3** (Covariant Difference of Metric Determinant). *The covariant finite difference of the metric determinant vanishes up to lattice contributions of order $\epsilon$, more precisely,*

$$D_a q = \epsilon \mu q_a \tag{4.4}$$

*where $\mu q_a$ is given by*

$$\mu q_a := \frac{1}{n!} \left( \mathcal{O}_a \left( q_{b_1 c_1}, \prod_{i=2}^{n} q_{b_i c_i} \right) + q_{b_1 c_1} \mathcal{O}_a \left( q_{b_2 c_2}, \prod_{i=3}^{n} q_{b_i c_i} \right) + \cdots + \prod_{i=1}^{n-2} q_{b_i c_i} \mathcal{O}_a(q_{b_{n-1} c_{n-1}}, q_{b_n c_n}) \right)$$
$$\cdot \varepsilon^{b_1 \ldots b_n} \varepsilon^{c_1 \ldots c_n}. \tag{4.5}$$

**Lemma 4.4** (Covariant Difference of Metric Determinant Square Root). *The covariant lattice difference of $\sqrt{q}$ vanishes up to lattice contributions of order $\epsilon$, more precisely,*

$$D_a \sqrt{q} = \epsilon \mu \sqrt{q}_a, \quad \text{where} \quad \mu \sqrt{q}_a := \frac{\Delta_a q \left( -\Delta_a^{f,\epsilon} \sqrt{q} + \Delta_a^{b,\epsilon} \sqrt{q} \right)}{2 \sqrt{q} \left( \sqrt{T_a q} + \sqrt{T_a^{-1} q} \right)} - \frac{\mu q_a}{2 \sqrt{q}}. \tag{4.6}$$

For later use, let us also compute the central lattice difference of $\sqrt{q}^{-1}$.

**Lemma 4.5** (Lattice Difference of $\sqrt{q}^{-1}$). *The (central) lattice difference of $\sqrt{q}^{-1}$ is given by*

$$\Delta_a \frac{1}{\sqrt{q}} = -\frac{\Delta_a q}{2 \sqrt{q}^3} + \epsilon \mu \sqrt{q}_a^{-1} \tag{4.7}$$

*where*

$$\mu \sqrt{q}_a^{-1} := \sqrt{(T_a^{-1} q)(T_a q)} \left( -\Delta_a^{f,\epsilon} \sqrt{q} + \Delta_a^{b,\epsilon} \sqrt{q} \right) + \sqrt{q(T_a q)} \Delta_a^{b,\epsilon} \sqrt{q} - q \Delta_a^{f,\epsilon} \sqrt{q}. \tag{4.8}$$

**Lemma 4.6** (Generalized Ricci Identity). *Let $T^{c_1 \ldots c_r}{}_{d_1 \ldots d_s}$ be an $(r,s)$–lattice tensor. Then the commutator of two covariant lattice differences acting upon $T$ is defined by*

$$[D_a, D_b] T^{c_1 \ldots c_r}{}_{d_1 \ldots d_s} := R^{c_1}{}_{fab} T^{fc_2 \ldots c_r}{}_{d_1 \ldots d_s} + \cdots + R^{c_r}{}_{fab} T^{c_1 \ldots c_{r-1} f}{}_{d_1 \ldots d_s}$$
$$- R^{f}{}_{d_1 ab} T^{c_1 \ldots c_r}{}_{fd_2 \ldots d_s} - \cdots - R^{f}{}_{d_s ab} T^{c_1 \ldots c_r}{}_{d_1 \ldots d_{s-1} f}$$

$$+ \epsilon \mu RI(T^{c_1 \ldots c_r}{}_{d_1 \ldots d_s})_{ab}, \tag{4.9}$$

where

$$\begin{aligned}
\mu RI(T^{c_1 \ldots c_r}{}_{d_1 \ldots d_s})_{ab} &= \left( \mathcal{O}_a(\Gamma^{c_1}{}_{fb}, T^{fc_2 \ldots c_r}{}_{d_1 \ldots d_s}) - \mathcal{O}_b(\Gamma^{c_1}{}_{fa}, T^{fc_2 \ldots c_r}{}_{d_1 \ldots d_s}) \right) + \ldots \\
&+ \left( \mathcal{O}_a(\Gamma^{c_r}{}_{fb}, T^{c_1 \ldots c_{r-1}f}{}_{d_1 \ldots d_s}) - \mathcal{O}_b(\Gamma^{c_r}{}_{fa}, T^{c_1 \ldots c_{r-1}f}{}_{d_1 \ldots d_s}) \right) \\
&- \left( \mathcal{O}_a(\Gamma^{f}{}_{d_1 b}, T^{c_1 \ldots c_r}{}_{fd_2 \ldots d_s}) - \mathcal{O}_b(\Gamma^{f}{}_{d_1 a}, T^{c_1 \ldots c_r}{}_{fd_2 \ldots d_s}) \right) - \ldots \\
&- \left( \mathcal{O}_a(\Gamma^{f}{}_{d_s b}, T^{c_1 \ldots c_r}{}_{d_1 \ldots d_{s-1}f}) - \mathcal{O}_b(\Gamma^{f}{}_{d_s a}, T^{c_1 \ldots c_r}{}_{d_1 \ldots d_{s-1}f}) \right).
\end{aligned} \tag{4.10}$$

**Lemma 4.7** (Riemann First Skew Symmetry). *The lattice Riemann tensor with lower indices $R_{abcd} := q_{af} R^{f}{}_{bcd}$ is skew symmetric up to errors of order $\epsilon$ with respect to permuting its anterior indices, i.e.,*

$$R_{abcd} = -R_{bacd} + \epsilon \mu R_{a \leftrightarrow bcd} \tag{4.11}$$

*where the lattice term is defined by*

$$\mu R_{a \leftrightarrow bcd} := \mathcal{O}_d(q_{af}, \Gamma^f{}_{cb}) - \mathcal{O}_c(q_{af}, \Gamma^f{}_{db}) + \mathcal{O}_d(q_{bf}, \Gamma^f{}_{ca}) - \mathcal{O}_c(q_{bf}, \Gamma^f{}_{da}). \tag{4.12}$$

**Lemma 4.8** (Riemann Second Skew Symmetry). *The lattice Riemann tensor with lower indices $R_{abcd}$ is skew symmetric with respect to permuting its posterior indices, i.e.,*

$$R_{abcd} = -R_{abdc}. \tag{4.13}$$

**Lemma 4.9** (Riemann Interchange Symmetry). *The lattice Riemann tensor with lower indices $R_{abcd}$ is symmetric upon interchanging the two pairs of anterior and posterior indices up to errors of order $\epsilon$, more precisely,*

$$R_{abcd} = R_{cdab} + \epsilon \mu R_{(ab), \leftrightarrow (cd)} \tag{4.14}$$

*where*

$$\mu R_{(ab) \leftrightarrow (cd)} := \mathcal{O}_d(q_{af}, \Gamma^f{}_{cb}) - \mathcal{O}_c(q_{af}, \Gamma^f{}_{db}) + \mathcal{O}_a(q_{cf}, \Gamma^f{}_{bd}) - \mathcal{O}_b(q_{cf}, \Gamma^f{}_{ad}). \tag{4.15}$$

**Lemma 4.10** (Jacobi Identity). *The Jacobi identity holds on the lattice for any lattice tensor T,*

$$([D_a, [D_b, D_c]] + [D_b, [D_c, D_a]] + [D_c, [D_a, D_b]])T = 0. \tag{4.16}$$

**Lemma 4.11** (First Bianchi Identity). *The first (algebraic) Bianchi identity holds on the lattice*

$$R_{abcd} + R_{acdb} + R_{adbc} = 0. \tag{4.17}$$

**Lemma 4.12** (Second Bianchi Identity). *The second Bianchi identity holds on the lattice up to lattice contributions proportional to $\epsilon$, in particular, it holds that*

$$D_a R^d{}_{fbc} + D_b R^d{}_{fca} + D_c R^d{}_{fab} = \epsilon \mu BI^d{}_{fabc}, \tag{4.18}$$

*where the lattice contribution is defined by*

$$\begin{aligned}
\mu BI^d{}_{fabc} &= \mathcal{O}_a(\Gamma^d{}_{cg}, \Gamma^g{}_{fb}) - \mathcal{O}_a(\Gamma^d{}_{bg}, \Gamma^g{}_{fc}) + \mathcal{O}_b(\Gamma^d{}_{ag}, \Gamma^g{}_{fc}) - \mathcal{O}_b(\Gamma^d{}_{cg}, \Gamma^g{}_{fa}) \\
&+ \mathcal{O}_c(\Gamma^d{}_{bg}, \Gamma^g{}_{fb}) - \mathcal{O}_c(\Gamma^d{}_{ag}, \Gamma^g{}_{fb}).
\end{aligned} \tag{4.19}$$

**Lemma 4.13** (Contracted Bianchi Identity). *The contraction of the second Bianchi identity on the lattice reads*

$$D_b R^b{}_a = \frac{1}{2} D_a R + \epsilon \, \mu CBI_a, \tag{4.20}$$

*where*

$$\mu CBI_a := \frac{1}{2}\Big(q^{fb} \mu BI^c{}_{fabc} + D_c(q^{fb} q^{cd} \mu R_{f \leftrightarrow dba}) - \mathcal{O}_a(q^{fb}, R_{fb}) - \mathcal{O}_b(q^{fb}, R_{fa}) - \mathcal{O}_c(q^{fb}, R^c{}_{fab})$$
$$-\mu D q_a{}^{fb} R_{fb} + \mu D q_b{}^{fb} R_{fa} + \mu D q_c{}^{fb} R^c{}_{fab}\Big) \tag{4.21}$$

Finally, let us introduce an operation which we denote by the lattice Lie difference.

**Definition 4.14** (Lattice Lie Difference). The Lie difference of a lattice tensor density $T$ of type $(r,s)$ and of weight $w$ in the direction of a lattice vector $X^a$ is given by

$$(\mathcal{L}_X T)^{b_1 \ldots b_r}{}_{c_1 \ldots c_s} := X^a \Delta^\epsilon_a T^{b_1 \ldots b_r}{}_{c_1 \ldots c_s} + w(\Delta_a X^a) T^{b_1 \ldots b_r}{}_{c_1 \ldots c_s} \tag{4.22}$$
$$- (\Delta_a X^{b_1}) T^{a b_2 \ldots b_r}{}_{c_1 \ldots c_s} - \cdots - (\Delta_a X^{b_r}) T^{b_1 \ldots b_{r-1} a}{}_{c_1 \ldots c_s}$$
$$+ (\Delta_{c_1} X^a) T^{b_1 \ldots b_r}{}_{a c_2 \ldots c_s} + \cdots + (\Delta_{c_s} X^a) T^{b_1 \ldots b_r}{}_{c_1 \ldots c_{s-1} a}.$$

## 5 Constraints and Constraint Algebra on the Lattice

With the notation and the results of the two previous sections, we reproduce the gravitational constraints on the lattice and compute the modified lattice Poisson algebra of the lattice constraints. The formerly continuous fields $q_{bc}(x)$ and $p^{de}(x)$ are promoted to a finite number of time–dependent variables $q^{X_1 \ldots X_n}_{bc}$ and $p^{de}_{X_1 \ldots X_n}$ at the spatial lattice sites $(X_1, \ldots, X_n) \in \mathbb{Z}^n$. Whenever possible, we suppress the lattice point indices to shorten notation.

It is convenient to split the Hamiltonian constraint into a kinetic and a potential part. On the lattice, we define

$$H_{\text{kin}}(N) := \epsilon^n \sum \frac{N}{\sqrt{q}} G_{abcd} p^{ab} p^{cd}, \qquad H_{\text{pot}}(N) := -\epsilon^n \sum N \sqrt{q} R \tag{5.1}$$

where summation is understood over all lattice points and all quantities appearing in the sum depend on the lattice points $X_1 \ldots X_n$. We define the supermetric $G_{abcd}$ by[1]

$$G_{abcd} := q_{ac} q_{bd} - \frac{1}{n-1} q_{ab} q_{cd}. \tag{5.2}$$

We should make clear what we mean by the lattice curvature scalar $R$ in $H^\epsilon_{\text{pot}}(N)$ because it can have different representations on the lattice that all reproduce the same continuum result (as do all lattice functions). Here, we choose to define the curvature scalar as follows:

$$R := q^{bd} R_{bd}, \tag{5.3}$$

---

[1] Note that DeWitt (1967) defines the supermetric differently but their contractions with the symmetric tensors $p^{ab} p^{cd}$ coincide.

$$R_{bd} := R^a{}_{bad}, \tag{5.4}$$

$$R^a{}_{bcd} := \Delta_c \Gamma^a{}_{db} - \Delta_d \Gamma^a{}_{cb} + \Gamma^a{}_{ce}\Gamma^e{}_{db} - \Gamma^a{}_{de}\Gamma^e{}_{cb}, \tag{5.5}$$

with the Christoffel symbols $\Gamma$ given in equation (4.1). With regards to the diffeomorphism constraints, we choose to promote the classical constraints as written in equation (2.6) to the following lattice constraints:

$$C(\mathcal{N}) := \epsilon^n \sum \mathcal{N}^a \left( -2\Delta_b(q_{ac}p^{cb}) + (\Delta_a q_{bc})p^{bc} \right). \tag{5.6}$$

In order to evaluate the constraint algebra, let us introduce the lattice Poisson bracket.

**Definition 5.1** (Lattice Poisson Bracket). For two functionals on the lattice $C : \mathcal{F}(\Lambda, \mathbb{R}) \to \mathbb{R}$ and $H : \mathcal{F}(\Lambda, \mathbb{R}) \to \mathbb{R}$ and two lattice functions $f, g \in \mathcal{F}(\Lambda, \mathbb{R})$ the lattice Poisson bracket is given by

$$\{C(f), H(g)\} = \frac{1}{\epsilon^n} \sum_{X_1\ldots X_n=0}^{N-1} \left( \frac{\partial C(f)}{\partial q_{ij}^{X_1\ldots X_n}} \frac{\partial H(g)}{\partial p^{ij}_{X_1\ldots X_n}} - \frac{\partial C(f)}{\partial p^{ij}_{X_1\ldots X_n}} \frac{\partial H(g)}{\partial q_{ij}^{X_1\ldots X_n}} \right). \tag{5.7}$$

Here, the lattice functionals $C(f)$ and $H(g)$ are to be understood as functions of the canonical variables $q_{ab}^{Y_1\ldots Y_n}$ and $p^{ab}_{Y_1\ldots Y_n}$ at each lattice site $Y_1 \ldots Y_n$ such that the derivatives are literally partial derivatives. Note that the factor $\epsilon^{-n}$ in this definition comes from properly transfering the Poisson bracket of continuous functionals to the lattice as we show in appendix B. In what follows, we will omit the lattice point indices.

### 5.1 $\{C(\mathcal{V}), C(\mathcal{W})\}$

In order to evaluate the Poisson bracket of two smeared diffeomorphism constraints, let us first express the diffeomorphism constraint in a different manner. By employing the technique of summation by parts (lemma 3.5), we can express it as

$$C(\mathcal{V}) = \epsilon^n \sum \left( 2(\Delta_b \mathcal{V}^a) q_{ac} p^{cb} - \Delta_a(\mathcal{V}^a p^{bc}) q_{bc} \right). \tag{5.8}$$

One can then easily show that the derivative of $C(\mathcal{V})$ with respect to $q_{ij}$ using $C(\mathcal{V})$ as given in equation (5.8) is

$$\frac{\partial C(\mathcal{V})}{\partial q_{ij}} = \epsilon^n \left( (S\Delta_a \mathcal{V}^i) p^{aj} + (\Delta_a \mathcal{V}^j) p^{ai} - \Delta_a(\mathcal{V}^a p^{ij}) \right), \tag{5.9}$$

see appendix C.1 for a proof. The same approach gives for the derivative of $C(\mathcal{V})$ with respect to $p^{ij}$

$$\frac{\partial C(\mathcal{V})}{\partial p^{ij}} = \epsilon^n \left( (\Delta_j \mathcal{V}^a) q_{ai} + (\Delta_i \mathcal{V}^a) q_{aj} + \mathcal{V}^a(\Delta_a q_{ij}) \right) = \epsilon^n (\mathcal{L}_\mathcal{V} q)_{ij}, \tag{5.10}$$

where we used definition 4.14 to write the result as the lattice Lie difference of $q_{ij}$. The Poisson bracket of two diffeomorphism constraints smeared with the lattice vectors $\mathcal{V}$ and $\mathcal{W}$ reads

$$\{C(\mathcal{V}), C(\mathcal{W})\} = \epsilon^{-n} \sum \left( \frac{\partial C(\mathcal{V})}{\partial q_{ij}} \frac{\partial C(\mathcal{W})}{\partial p^{ij}} - \frac{\partial C(\mathcal{V})}{\partial p^{ij}} \frac{\partial C(\mathcal{W})}{\partial q_{ij}} \right) \tag{5.11}$$

such that its evaluation simply requires to multiply equations (5.9) and (5.10) with the correct smearing lattice functions. Examplarily, let us consider the first part of the Poisson bracket. The rationale is to shift the Lie difference $\mathcal{L}_\mathcal{V}$ acting on $p^{ij}$ to $(\mathcal{L}_\mathcal{W} q)_{ij}$. Using summation by parts this gives (see appendix C.1)

$$\epsilon^{-n} \sum \frac{\partial C(\mathcal{V})}{\partial q_{ij}} \frac{\partial C(\mathcal{W})}{\partial p^{ij}} = \epsilon^n \sum \left( \mathcal{L}_\mathcal{V}\left(p^{ij}(\mathcal{L}_\mathcal{W} q)_{ij}\right) - p^{ij}(\mathcal{L}_\mathcal{V}(\mathcal{L}_\mathcal{W} q))_{ij} \right). \tag{5.12}$$

The first contribution is the Lie difference of a scalar density of weight one and hence vanishes due to our choice of periodic boundary conditions

$$\sum \mathcal{L}_\mathcal{V}\left(p^{ij}(\mathcal{L}_\mathcal{W} q)_{ij}\right) = \sum \Delta_a \left(\mathcal{V}^a p^{ij}(\mathcal{L}_\mathcal{W} q)_{ij}\right) = 0. \tag{5.13}$$

We consequently get for the Poisson bracket

$$\{C(\mathcal{V}), C(\mathcal{W})\} = \epsilon^n \sum p^{ij}\left((\mathcal{L}_\mathcal{W}(\mathcal{L}_\mathcal{V} q))_{ij} - (\mathcal{L}_\mathcal{V}(\mathcal{L}_\mathcal{W} q))_{ij}\right). \tag{5.14}$$

In the continuum, this expression would reduce to $C(\mathcal{L}_\mathcal{V}(\mathcal{W}))$ because of the properties of the Lie derivate. On the lattice however, the result includes additional lattice terms that scale with $\epsilon$. As we show in appendix C.1, we obtain by a straightforward evaluation of the Lie differences and by applying the product rule that

$$\{C(\mathcal{V}), C(\mathcal{W})\} = C(\mathcal{L}_\mathcal{V}(\mathcal{W})) + \epsilon A_{CC}(\mathcal{V}, \mathcal{W}) \tag{5.15}$$

where the Lie difference of $\mathcal{W}$ with respect to $\mathcal{V}$ is given by

$$\mathcal{L}_\mathcal{V}(\mathcal{W})^b = \mathcal{V}^a(\Delta_a \mathcal{W}^b) - \mathcal{W}^a(\Delta_a \mathcal{V}^b), \tag{5.16}$$

and the additonal lattice term reads

$$A_{CC}(\mathcal{V}, \mathcal{W}) = \epsilon^n \sum 2 p^i{}_b \left( \mathcal{O}_i(\mathcal{W}^a, \Delta_a \mathcal{V}^b) - \mathcal{O}_i(\mathcal{V}^a, \Delta_a \mathcal{W}^b) \right) + B(\mathcal{V}, \mathcal{W}) - B(\mathcal{W}, \mathcal{V}), \tag{5.17}$$

where we used the abbreviation

$$B(\mathcal{V}, \mathcal{W}) = \epsilon^n \sum p^{ij}\left(\mathcal{V}^a\left(2\mathcal{O}_a(\Delta_i \mathcal{W}^b, q_{bj}) + \mathcal{O}_a(\mathcal{W}^b, \Delta_b q_{ij})\right)\right) \tag{5.18}$$

with $p^i{}_b := p^{ic} q_{cb}$. The above outcome corresponds to the continuum result up to the lattice term $\epsilon A_{CC}$.

## 5.2 $\{H(f), H(g)\}$

The Poisson bracket of two smeared Hamiltonian constraints simplifies by splitting $H(f)$ into its kinetic and potential parts

$$\{H(f), H(g)\} = \{H_{\text{kin}}(f) + H_{\text{pot}}(f), H_{\text{kin}}(g) + H_{\text{pot}}(g)\}$$
$$= \{H_{\text{kin}}(f), H_{\text{kin}}(g)\} + \{H_{\text{pot}}(f), H_{\text{kin}}(g)\} + \{H_{\text{kin}}(f), H_{\text{pot}}(g)\}, \quad (5.19)$$

where we used that $H_{\text{pot}}(f)$ does not depend on $p^{ij}$ such that the Poisson bracket $\{H_{\text{pot}}(f), H_{\text{pot}}(g)\}$ vanishes identically.

We begin by considering the first contribution in equation (5.19). For this, it suffices to recall that $H_{\text{kin}}(f)$ does not include any difference operators and depends on products of $q_{ij}$ and $p^{ij}$ with $f$. As a consequence, its derivatives with respect to $q_{ij}$ and $p^{ij}$ respectively factorize the smearing field $f$. This motivates us to define the Hamiltonian lattice densities

$$\mathcal{H}_{\text{kin}} := \frac{1}{\sqrt{q}} G_{abcd} p^{ab} p^{cd}, \quad \mathcal{H}_{\text{pot}} := -\sqrt{q} R. \quad (5.20)$$

Note that similar to the continuum case, densities are functions (not functionals) of the canonical lattice functions $q_{ab}, p^{ab} \in \mathcal{F}(\Lambda, \mathbb{R})$. With this, we obtain

$$\frac{\partial H_{\text{kin}}(f)}{\partial q_{ij}} = \epsilon^n f \frac{\partial \mathcal{H}_{\text{kin}}}{\partial q_{ij}}, \quad \frac{\partial H_{\text{pot}}(f)}{\partial q_{ij}} = \epsilon^n f \frac{\partial \mathcal{H}_{\text{pot}}}{\partial q_{ij}}. \quad (5.21)$$

Hence, the Poisson bracket of $H_{\text{kin}}(f)$ and $H_{\text{kin}}(g)$ reduces to

$$\{H_{\text{kin}}(f), H_{\text{kin}}(g)\} = \epsilon^n \sum (fg - gf) \frac{\partial \mathcal{H}_{\text{kin}}}{\partial q_{ij}} \frac{\partial \mathcal{H}_{\text{kin}}}{\partial p^{ij}} = 0. \quad (5.22)$$

We manipulate the two remaining contributions to the Poisson bracket in equation (5.19) using that $H_{\text{pot}}(f)$ does not depend on $p^{ij}$, and get

$$\{H_{\text{pot}}(f), H_{\text{kin}}(g)\} + \{H_{\text{kin}}(f), H_{\text{pot}}(g)\} = \epsilon^{-n} \sum \left( \frac{\partial H_{\text{pot}}(f)}{\partial q_{ij}} \frac{\partial H_{\text{kin}}(g)}{\partial p^{ij}} - \frac{\partial H_{\text{kin}}(f)}{\partial p^{ij}} \frac{\partial H_{\text{pot}}(g)}{\partial q_{ij}} \right). \quad (5.23)$$

In order to evaluate this expression, we need the derivative of $H_{\text{pot}}(f)$ with respect to $q_{ij}$. We use that by definition $R = q^{ab} R_{ab}$ and employ the product rule to get

$$\frac{\partial H_{\text{pot}}(f)}{\partial q_{ij}} = -\epsilon^n \frac{\partial}{\partial q_{ij}} \sum f \sqrt{q} q^{ab} R_{ab} = \epsilon^n f \sqrt{q} (G^{ij} + \epsilon C^{ij}) - \epsilon^n \sum f \sqrt{q} q^{ab} \frac{\partial R_{ab}}{\partial q_{ij}}, \quad (5.24)$$

where we defined the lattice Einstein tensor $G^{ij} := R^{ij} - q^{ij} R/2$ in analogy to the continuum case and the additional lattice contribution by

$$C^{ij} = \frac{1}{2} q^{ja} q^{ib} q^{cd} \mu R_{(da) \leftrightarrow (cb)}, \quad (5.25)$$

see appendix C.2 for a proof of this identity. Since the first term in equation (5.24) is proportional to $f$, and $\partial H_{\text{kin}}(g)/\partial p^{ij}$ is proportional to $g$, their product will simply cancel with the second contribution to the Poisson bracket in equation (5.23). This follows from applying the same rationale as for the purely kinetic contribution to the Poisson bracket. Hence, we can ignore this term.

The second term in equation (5.24) which includes the derivative of the lattice Ricci tensor with respect to $q_{ij}$ requires more care. Therefore, let us first note that the partial derivative with respect to $q_{ij}$ and the lattice finite difference operator $\Delta_a$ applied to any lattice function $f \in \mathcal{F}(\Lambda, \mathbb{R})$ commute (see lemma C.1 for a proof). Let then $\partial f$ be a short hand notation for the derivative of $f$ with respect to the metric tensor. It is then straightforward to establish that

$$\partial R_{ab} = \frac{1}{2}\big(D_c(q^{cd}D_a\,\partial q_{db}) + D_c(q^{cd}D_b\,\partial q_{da}) - D_a(q^{cd}D_b\,\partial q_{cd}) - D_c(q^{cd}D_d\,\partial q_{ab})\big), \qquad (5.26)$$

see lemmas C.2 and C.3 ff. When plugged into the last expression in equation (5.24), it is useful to have the covariant differences in $\partial R_{ab}$ act exclusively on the smearing lattice function $f$. Using first summation by parts and then evaluating the covariant differences such that they act only upon $f$, we find that

$$-\epsilon^n \sum f \sqrt{q}\, q^{ab} \frac{\partial R_{ab}}{\partial q_{ij}} = \epsilon^n \sqrt{q}\,\big(q^{ij}q^{ab} - q^{ai}q^{bj}\big) D_a(D_b f) + \epsilon\, E^{ij}(f) \qquad (5.27)$$

where the additional lattice contribution $E^{ij}(f)$ reads

$$E^{ij}(f) = \frac{1}{2}\big(F^{ij}(f, q^{ij}, q^{ab}) + F^{ij}(f, q^{ab}, q^{ij}) - F^{ij}(f, q^{ai}, q^{bj}) - F^{ij}(f, q^{bj}, q^{ai})\big) \qquad (5.28)$$

for which we defined $F^{ij}$ in order to distinguish the lattice terms for each contribution to $\partial R_{ab}$,

$$F^{ij}(f, q^{ij}, q^{ab}) := \epsilon^{n-1}\big((D_b f)D_a(\sqrt{q}q^{ij}q^{ab}) + D_a(fq^{ab}D_b(\sqrt{q}q^{ij}))\big) + \epsilon^n \mathcal{O}_b(D_a f, \sqrt{q}q^{ij}q^{ab})$$
$$+ \epsilon^n D_a(q^{ab}\mathcal{O}_b(f, \sqrt{q}q^{ij})). \qquad (5.29)$$

We refer again to appendix C.2 for a more detailed derivation of these results. Note that while the first two terms formally enter with a factor $\epsilon^{-1}$, the covariant derivatives of the determinant of $q$ as well as of the inverse metric are proportional to $\epsilon$ such that $\epsilon^{-1}$ cancels (see lemmas 4.2 and 4.3).

In a final step, we multiply equation (5.27) with

$$\frac{\partial H_{\text{kin}}(g)}{\partial p^{ij}} = \epsilon^n \frac{2g}{\sqrt{q}}\Big(p_{ij} - \frac{p}{n-1}q_{ij}\Big) \qquad (5.30)$$

and subtract the same product with $g$ and $f$ interchanged in order to compute the total Poisson bracket. For the first part, we get

$$\epsilon^n \sum 2(f D_a D_b g - g D_a D_b f) p^{ab}. \qquad (5.31)$$

Using summation by parts and applying the product rule, this is equivalent to

$$\epsilon^n \sum q^{ad}(f(\Delta_d g) - g(\Delta_d f))(-2\Delta_b(q_{ac}p^{cb}) + (\Delta_a q_{bc})p^{bc}) + \epsilon I(f,g), \quad (5.32)$$

where the lattice term $I(f,g)$ is given by

$$I(f,g) = \epsilon^n \sum 2(\mathcal{O}_b(g, p^{db})\Delta_d f - \mathcal{O}_b(f, p^{db})\Delta_d g + \mathcal{O}_b(q_{ad}, p^{db})q^{ac}(f\Delta_c g - g\Delta_c f)). \quad (5.33)$$

In equation (5.32), we identify the diffeomorphism constraint smeared with the lattice vector

$$\mathcal{V}^a(q, f, g) := q^{ad}(f\Delta_d g - g\Delta_d f). \quad (5.34)$$

Finally, we add the lattice term $\epsilon E^{ij}$ from equation (5.27) and get

$$\{H(f), H(g)\} = C(\mathcal{V}) + \epsilon A_{HH}(f,g), \quad (5.35)$$

where the additional lattice contribution is given by

$$A_{HH}(f,g) = \epsilon^n \sum \frac{2}{\sqrt{q}}\left(p_{ij} - \frac{p}{n-1}q_{ij}\right)(gE^{ij}(f) - fE^{ij}(g)) + I(f,g). \quad (5.36)$$

This corresponds to the continuum result plus the additional lattice contribution $\epsilon A_{HH}$.

## 5.3 $\{C(\mathcal{V}), H(f)\}$

The Poisson bracket of a smeared diffeomorphism constraint and a smeared Hamiltonian constraint reads

$$\{C(\mathcal{V}), H(f)\} = \epsilon^{-n} \sum \left(\frac{\partial C(\mathcal{V})}{\partial q_{ij}}\frac{\partial H(f)}{\partial p^{ij}} - \frac{\partial C(\mathcal{V})}{\partial p^{ij}}\frac{\partial H(f)}{\partial q_{ij}}\right) \quad (5.37)$$

for a smearing lattice function $f$ and a lattice vector $\mathcal{V}$. In order to compute the first contribution to the Poisson bracket, we use that

$$\frac{\partial C(\mathcal{V})}{\partial q_{ij}} = \epsilon^n((\Delta_a \mathcal{V}^i)p^{aj} + (\Delta_a \mathcal{V}^j)p^{ai} - \Delta_a(\mathcal{V}^a p^{ij})), \quad (5.38)$$

$$\frac{\partial H(f)}{\partial p^{ij}} = \epsilon^n\left(\frac{2f}{\sqrt{q}}\left(p_{ij} - \frac{p}{n-1}q_{ij}\right)\right). \quad (5.39)$$

We multiply those expressions and apply the product rule to the last contribution to equation (5.38) (see appendix C.3 for more details). With regards to the second contribution to the Poisson bracket, let us first focus on only the kinetic part of the Hamilton constraint contributing to this term. With this, we can express the respective partial derivatives as follows:

$$\frac{\partial C(\mathcal{V})}{\partial p^{ij}} = \epsilon^n((\Delta_j \mathcal{V}^a)q_{ai} + (\Delta_i \mathcal{V}^a)q_{aj} + \mathcal{V}^a(\Delta_a q_{ij})), \quad (5.40)$$

$$\frac{\partial H_{\text{kin}}(f)}{\partial q_{ij}} = -\epsilon^n \frac{f}{2} q^{ij} \mathcal{H}_{\text{kin}} + \epsilon^n \frac{2f}{\sqrt{q}} \left( p^i{}_b p^{bj} - \frac{p}{n-1} p^{ij} \right). \tag{5.41}$$

We perform a straightforward multiplication of the two expressions. By comparing with the result for the first contribution to the bracket, we find that several terms cancel and simplify. We are left with

$$\epsilon^{-n} \sum \left( \frac{\partial C(\mathcal{V})}{\partial q_{ij}} \frac{\partial H(f)}{\partial p^{ij}} - \frac{\partial C(\mathcal{V})}{\partial p^{ij}} \frac{\partial H_{\text{kin}}(f)}{\partial q_{ij}} \right) = \epsilon^n \sum \mathcal{V}^d (\Delta_d f) \mathcal{H}_{\text{kin}} + \epsilon J(f, \mathcal{V}), \tag{5.42}$$

where the additional lattice term $J(f, \mathcal{V})$ is defined by

$$J(f, \mathcal{V}) := \epsilon^n \sum \left( f \mathcal{O}_d(\mathcal{V}^d, \mathcal{H}_{\text{kin}}) + \mathcal{O}_d(\mathcal{V}^d, p^{ij}) f \frac{\partial \mathcal{H}_{\text{kin}}}{\partial p^{ij}} - f \mathcal{V}^d K_d \right). \tag{5.43}$$

$K_d$ in this expression comes from using a "lattice chain rule" which establishes the following identity

$$(\Delta_d p^{ij}) \frac{\partial \mathcal{H}_{\text{kin}}}{\partial p^{ij}} + (\Delta_d q_{ij}) \frac{\partial \mathcal{H}_{\text{kin}}}{\partial q_{ij}} = \Delta_d \mathcal{H}_{\text{kin}} + \epsilon K_d, \tag{5.44}$$

The lattice term $K_d$ is accordingly given by

$$K_d = \left( -\frac{\mu q_d}{2\sqrt{q}^3} + \mu \sqrt{q}_d^{-1} \right) \sqrt{q} \mathcal{H}_{\text{kin}} + \frac{p^{ab} p^{ce}}{\sqrt{q}} \left( \mathcal{O}_d(q_{ac}, q_{be}) - \frac{\mathcal{O}_d(q_{ab}, q_{ce})}{n-1} \right) + \mathcal{O}_d \left( \frac{1}{\sqrt{q}}, \sqrt{q} \mathcal{H}_{\text{kin}} \right)$$
$$+ \frac{1}{\sqrt{q}} \mathcal{O}_d(G_{abce}, p^{ab} p^{ce}) + \frac{G_{abce}}{\sqrt{q}} \mathcal{O}_d(p^{ab}, p^{ce}). \tag{5.45}$$

see lemma C.4 for a proof. After using this chain rule, we used summation by parts in order to shift the difference operator onto $f$ in equation (5.42).

The only remaining contribution to the Poisson bracket involves on the one hand the expression

$$\frac{\partial H_{\text{pot}}(f)}{\partial q_{ij}} = \epsilon^n f \sqrt{q} (G^{ij} + \epsilon C^{ij}) + \epsilon^n \sqrt{q} (q^{ij} q^{ab} - q^{ai} q^{bj}) D_a(D_b f) + \epsilon E^{ij}(f) \tag{5.46}$$

which we know from the previous section and where the lattice terms $C^{ij}(f)$ and $E^{ij}$ are respectively defined in equations (5.25) and (5.28). On the other hand, it involves the derivative of $C(\mathcal{V})$ with respect to $p^{ij}$ that we write in a slightly different manner than before as

$$\frac{\partial C(\mathcal{V})}{\partial p^{ij}} = \epsilon^n \big( (D_j \mathcal{V}^a) q_{ai} + (D_i \mathcal{V}^a) q_{aj} \big). \tag{5.47}$$

We multiply these two expressions and start by defining the lattice terms proportional to $\epsilon$ as

$$\epsilon M(f, \mathcal{V}) := -\epsilon^{n+1} \sum \big( (D_j \mathcal{V}^a) q_{ai} + (D_i \mathcal{V}^a) q_{aj} \big) \big( f \sqrt{q} C^{ij} + E^{ij}(f) \big). \tag{5.48}$$

For the term proportional to the Einstein tensor, we get

$$\epsilon^n \sum \sqrt{q} \mathcal{V}^a \left( 2R^b{}_a \Delta_b f - R \Delta_a f \right) + \epsilon N(f, \mathcal{V}) \tag{5.49}$$

where the additional lattice contributions $N(f, \mathcal{V})$ are given by

$$N(f, \mathcal{V}) = \epsilon^n \sum \left( f \sqrt{q} q^{jb} q^{cd} \mu R_{(ca)\leftrightarrow(db)} + 2f \mathcal{V}^a \mu D \sqrt{q}_b R^b{}_a - f \mathcal{V}^a R \mu D \sqrt{q}_a \right). \tag{5.50}$$

This lattice term comes from interchanging indices of the Riemann tensor, see lemma 4.9 for the definition of $\mu R_{(da)\leftrightarrow(cb)}$, and from shifting the covariant differences from $\mathcal{V}$ to $f$. Finally, the last step consists in multiplying the second contribution to $\partial \mathcal{H}_{\text{pot}}(f)/\partial q_{ij}$ with $\partial C(\mathcal{V})/\partial p^{ij}$, i.e., to evaluate

$$-\sum \left( (D_j \mathcal{V}^a) q_{ai} + (D_i \mathcal{V}^a) q_{aj} \right) \sqrt{q} \left( q^{ij} q^{ab} - q^{ai} q^{bj} \right) D_a(D_b f). \tag{5.51}$$

Once more, it is useful to shift the covariant difference operators from $\mathcal{V}$ to $f$ in this expression and then to apply the generalized Ricci identity, see appendix C.3 for more details. This yields

$$-\epsilon^n \sum 2 \sqrt{q} \mathcal{V}^a R^b{}_a \Delta_b f + \epsilon S(f, \mathcal{V}), \tag{5.52}$$

where $S(f, \mathcal{V})$ is defined by

$$S(f, \mathcal{V}) := \epsilon^n \sum 2 \sqrt{q} \mathcal{V}^c q^{ab} \left( q^{df} \mu R_{f \leftrightarrow bac} \Delta_d f - \mu RI(D_b f)_{ac} \right) + T(f, \mathcal{V}) \tag{5.53}$$

with

$$T(f, \mathcal{V}) := \epsilon^n \sum 2 \mathcal{V}^c \left( \mu D \sqrt{q}_c D^a D_a f - \mu D \sqrt{q}_a q^{ba} D_c D_b f + \sqrt{q} \mu D q_c{}^{ab} D_a D_b f - \sqrt{q} \mu D q_a{}^{ab} D_c D_b f \right.$$
$$\left. + \sqrt{q} \left( \mathcal{O}_c(q^{ab}, D_a D_b f) - \mathcal{O}_a(q^{ab}, D_c D_b f) \right) + \mathcal{O}_c(\sqrt{q}, D^a D_a f) - \mathcal{O}_a(\sqrt{q}, q^{ab} D_c D_b f) \right). \tag{5.54}$$

We are ready to add all contributions, in particular equations (5.42), (5.48), (5.49) and (5.52). Two terms cancel and we end up with

$$\{C(\mathcal{V}), H(f)\} = \epsilon^n \sum \mathcal{V}^d (\Delta_d f) \left( \mathcal{H}_{\text{kin}} - \sqrt{q} R \right) + \epsilon A_{CH}(f, \mathcal{V}) \tag{5.55}$$

where we defined

$$A_{CH}(f, \mathcal{V}) := J(f, \mathcal{V}) + M(f, \mathcal{V}) + N(f, \mathcal{V}) + S(f, \mathcal{V}). \tag{5.56}$$

Knowing that $\mathcal{H}_{\text{pot}} = -\sqrt{q} R$ and with $\mathcal{H} := \mathcal{H}_{\text{kin}} + \mathcal{H}_{\text{pot}}$, the final result is

$$\{C(\mathcal{V}), H(f)\} = \epsilon^n \sum \mathcal{V}^d (\Delta_d f) \mathcal{H} + \epsilon A_{CH}(f, \mathcal{V})$$
$$= H(\mathcal{L}_\mathcal{V} f) + \epsilon A_{CH}(f, \mathcal{V}) \tag{5.57}$$

where we employed that $\mathcal{L}_\mathcal{V} f = \mathcal{V}^d \Delta_d f$.

## 6   Conclusion

In this series of papers, of which this is the second one, we aim to overcome the longstanding obstacles that have hindered the progress of quantum geometrodynamics by proposing novel solutions.

Our first step towards achieving this goal consists in regularizing the classical Hamiltonian approach to gravity. Specifically, we reduce its phase space to the set of piecewise constant functions on the spatial manifold. With this, our approach amounts to confining the theory to a regular lattice with lattice constant $\epsilon$. We are then able to construct a discrete theory of gravity with the tools of conventional lattice theory. Besides, we implement periodic boundary conditions which allow us to study an infinite spacetime within a finite region. We represent the constraints on the lattice and compute their Poisson algebra. It is not surprising that the hypersurface deformation algebra does not close anymore as we put gravity on the lattice. In fact, we obtain anomalies that are proportional to the lattice constant, and which depend on the canonical variables at the respective lattice sites. These functions are regular on the phase space of positive definite metrics, and mix the variables of different lattice sites.

We note that the restriction of the phase space to piecewise constant functions required the modification of the constraint algebra in the sense that derivatives were replaced by finite differences. This resembles the finite differences method in the theory of numerical partial differential equations and is akin to what is being done in numerical relativity. Another popular method would be finite element analysis. There, it is possible to use higher order approximations and, e.g. express the spatial metric in terms of piecewise quadratic functions. One is then freed from the necessity to replace derivatives by finite differences and the lattice corrections would vanish identically. The price we would pay for this is that we would get additional degrees of freedom on each lattice site, corresponding to the additional slopes needed to parametrize the piecewise quadratic functions and these additional degrees of freedom would also have to be quantized later on. We will explore this possibility in a future publication.

Our findings serve as the starting point for constructing a well–defined quantum theory of gravity in its ADM incarnation. Since our reduction scheme confines classical gravity to a finite number of degrees of freedom, we can define a Hilbert space at each lattice site following the rules of standard quantum mechanics. More precisely, we introduce a quantum representation of the standard commutation relations of the spatial metric and its conjugate momenta at each lattice site that is able to realize positive definiteness of the metric tensor (Lang and Schander 2023a). With a simple tensor product, we combine all lattice site Hilbert spaces to a total Hilbert space. With this procedure, we have constructed, for the first time, a well–defined Hilbert space for quantum geometrodynamics on the lattice. In upcoming publications, we address the issue of taking the continuum limit of our proposed lattice quantum theory.

*Acknowledgements*

We would like to express our gratitude to Lee Smolin and to Renate Loll for their helpful comments on this project. We also extend our sincere appreciation to Bianca Dittrich for insightful discussions. Our special thanks go to Thomas Thiemann for detailed discussions during the final phase of our work, as well as for the hospitality of the Insitute for Quantum Gravity at FAU Erlangen–Nürnberg.

Research at Perimeter Institute is supported in part by the Government of Canada through the Department of Innovation, Science and Economic Development and by the Province of Ontario through the Ministry of Colleges and Universities.

## A   Formulas in Riemannian Lattice Geometry

In this appendix, we provide more details on the derivation of the formulas in Riemannian lattice geometry as presented in section 4. In particular, we state the lemmas again with their respective proofs.

**Lemma A.1** (Covariant difference of inverse metric). *The covariant difference of the inverse metric vanishes up to lattice terms, i.e.,*

$$D_a q^{bc} = \epsilon \mu D q_a{}^{bc}, \quad \text{where} \quad \mu D q_a{}^{bc} := -q^{bd} \mathcal{O}_a(q_{de}, q^{ec}). \tag{A.1}$$

*Proof.* By definition the metric and its inverse satisfy $q_{de} q^{ec} = \delta_d^c$, where $\delta_d^c$ is the Kronecker delta. Applying the covariant lattice difference to this equation gives

$$\begin{aligned} 0 = D_a(q_{de} q^{ec}) &= (D_a q_{de}) q^{ec} + q_{de}(D_a q^{ec}) + \epsilon \, \mathcal{O}_a(q_{de}, q^{ec}) \\ &= q_{de}(D_a q^{ec}) + \epsilon \, \mathcal{O}_a(q_{de}, q^{ec}) \end{aligned} \tag{A.2}$$

where we used that $D_a q_{de} = 0$ as can easily seen by evaluating the Christoffel symbols in this expression. Contraction with $q^{bd}$ yields

$$D_a q^{bc} = -\epsilon \, q^{bd} \mathcal{O}_a(q_{de}, q^{ec}). \tag{A.3}$$

□

**Lemma A.2** (Covariant Difference of Metric Determinant). *The covariant finite difference of the metric determinant vanishes up to lattice contributions of order $\epsilon$, more precisely,*

$$D_a q = \epsilon \mu q_a \tag{A.4}$$

*where $\mu q_a$ is given in equation (A.7).*

*Proof.* We use the definition of the finite covariant difference of a lattice function of density weight two, namely,

$$D_a q = \Delta_a q - 2 \Gamma^b{}_{ba} q. \tag{A.5}$$

To compute the finite difference $\Delta_a$ of $q$, we exploit the representation of $q$ in terms of the metric and the Levi–Civita symbol $\varepsilon^{b_1 \ldots b_n}$ in $n$ dimensions as well as the lattice product rule and obtain

$$\Delta_a q = \frac{1}{n!} \varepsilon^{b_1 \ldots b_n} \varepsilon^{c_1 \ldots c_n} \Delta_a \prod_{i=1}^n q_{b_i c_i} \tag{A.6}$$

$$= \frac{1}{n!} \varepsilon^{b_1\ldots b_n} \varepsilon^{c_1\ldots c_n} \left( (\Delta_a q_{b_1 c_1}) \prod_{i=2}^{n} q_{b_i c_i} + q_{b_1 c_1}(\Delta_a q_{b_2 c_2}) \prod_{i=3}^{n} q_{b_i c_i} + \cdots + \prod_{i=1}^{n-1} q_{b_i c_i}(\Delta_a q_{b_n c_n}) \right)$$

$$+ \varepsilon \left( \mathcal{O}_a\!\left(q_{b_1 c_1}, \prod_{i=2}^{n} q_{b_i c_i}\right) + q_{b_1 c_1} \mathcal{O}_a\!\left(q_{b_2 c_2}, \prod_{i=3}^{n} q_{b_i c_i}\right) + \cdots + \prod_{i=1}^{n-2} q_{b_i c_i} \mathcal{O}_a(q_{b_{n-1} c_{n-1}}, q_{b_n c_n}) \right)$$

$$\cdot \frac{1}{n!} \varepsilon^{b_1\ldots b_n} \varepsilon^{c_1\ldots c_n}.$$

By relabeling indices the sum in the first line reduces to $n$ times the same term while we define the lattice term proportional to $\varepsilon$ in the second and third line to be

$$\mu q_a := \frac{1}{n!} \left( \mathcal{O}_a\!\left(q_{b_1 c_1}, \prod_{i=2}^{n} q_{b_i c_i}\right) + q_{b_1 c_1} \mathcal{O}_a\!\left(q_{b_2 c_2}, \prod_{i=3}^{n} q_{b_i c_i}\right) + \cdots + \prod_{i=1}^{n-2} q_{b_i c_i} \mathcal{O}_a(q_{b_{n-1} c_{n-1}}, q_{b_n c_n}) \right)$$
$$\cdot \varepsilon^{b_1\ldots b_n} \varepsilon^{c_1\ldots c_n}. \tag{A.7}$$

This gives in total

$$\Delta_a q = \frac{1}{(n-1)!} \varepsilon^{b_1\ldots b_n} \varepsilon^{c_1\ldots c_n} \prod_{i=2}^{n} q_{b_i c_i}(\Delta_a q_{b_1 c_1}) + \varepsilon \mu q_a = q\, q^{b_1 c_1}(\Delta_a q_{b_1 c_1}) + \varepsilon \mu q_a, \tag{A.8}$$

where we inserted an explicit formula for the inverse metric in terms of the metric. Let us consider the second contribution

$$-2\Gamma^{b}{}_{ba} q = -q^{bc}(\Delta_b q_{ca} + \Delta_a q_{bc} - \Delta_c q_{ba})\, q. \tag{A.9}$$

Since $q^{bc}$ is symmetric, the first and the second contribution on the right hand side of this expression cancel and we are left with

$$-2\Gamma^{b}{}_{ba} = -q\, q^{bc} \Delta_a q_{bc}. \tag{A.10}$$

Adding equation (A.8) and equation (A.9) gives the desired result. □

**Lemma A.3** (Covariant Difference of Metric Determinant Square Root)**.** *The covariant lattice difference of $\sqrt{q}$ vanishes up to lattice contributions of order $\varepsilon$, more precisely*

$$D_a \sqrt{q} = \varepsilon \mu \sqrt{q}_a, \tag{A.11}$$

*where $\mu\sqrt{q}_a$ is defined by equation* (A.13)*.*

*Proof.* First, we use the expression of the central difference and the finite covariant difference in order to write

$$D_a \sqrt{q} = \left( \Delta_a \sqrt{q} - \sqrt{q}\, \Gamma^{b}{}_{ba} \right) \tag{A.12}$$

$$= \frac{\sqrt{T_a q} - \sqrt{T_a^{-1} q}}{\varepsilon} - \frac{\sqrt{q}}{2} q^{bc} \Delta_a q_{bc}$$

$$= \frac{\Delta_a q}{\sqrt{T_a q} + \sqrt{T_a^{-1} q}} - \frac{\Delta_a q - \epsilon \mu q_a}{2\sqrt{q}}$$

$$= \frac{\Delta_a q \left( \sqrt{q} - \frac{1}{2}\left(\sqrt{T_a q} + \sqrt{T_a^{-1} q}\right)\right)}{\sqrt{q}\left(\sqrt{T_a q} + \sqrt{T_a^{-1} q}\right)} - \frac{\epsilon \mu q_a}{2\sqrt{q}}$$

$$= \frac{\epsilon \Delta_a q \left(-\Delta_a^{\mathrm{f}} \sqrt{q} + \Delta_a^{\mathrm{b}} \sqrt{q}\right)}{2\sqrt{q}\left(\sqrt{T_a q} + \sqrt{T_a^{-1} q}\right)} - \frac{\epsilon \mu q_a}{2\sqrt{q}}.$$

Then, we define

$$\mu \sqrt{q}_a := \frac{\Delta_a q \left(-\Delta_a^{\mathrm{f}} \sqrt{q} + \Delta_a^{\mathrm{b}} \sqrt{q}\right)}{2\sqrt{q}\left(\sqrt{T_a q} + \sqrt{T_a^{-1} q}\right)} - \frac{\mu q_a}{2\sqrt{q}} \tag{A.13}$$

which completes the proof. $\square$

**Lemma A.4** (Lattice Difference of $\sqrt{q}^{-1}$). *The (central) lattice difference of $\sqrt{q}^{-1}$ is given by*

$$\Delta_a \frac{1}{\sqrt{q}} = -\frac{\Delta_a q}{2\sqrt{q}^3} + \epsilon \mu \sqrt{q}_a^{-1}, \tag{A.14}$$

*where*

$$\mu \sqrt{q}_a^{-1} = \sqrt{(T_a^{-1} q)(T_a q)}\left(-\Delta_a^{f} \sqrt{q} + \Delta_a^{b} \sqrt{q}\right) + \sqrt{q(T_a q)}\, \Delta_a^{b} \sqrt{q} - q\, \Delta_a^{f} \sqrt{q}. \tag{A.15}$$

*Proof.* By definition, the covariant difference of $\sqrt{q}^{-1}$ on the lattice is given by

$$\Delta_a \frac{1}{\sqrt{q}} = \frac{1}{\epsilon}\left(\frac{1}{\sqrt{T_a q}} - \frac{1}{\sqrt{T_a^{-1} q}}\right). \tag{A.16}$$

Bringing both terms to the same denominator and using the binomial theorem, we get

$$\Delta_a \frac{1}{\sqrt{q}} = \frac{1}{\epsilon}\left(-\frac{T_a q - T_a^{-1} q}{\sqrt{(T_a q)(T_a^{-1} q)}\left(\sqrt{T_a q} + \sqrt{T_a^{-1} q}\right)}\right) = -\frac{\Delta_a q}{\sqrt{(T_a q)(T_a^{-1} q)}\left(\sqrt{T_a q} + \sqrt{T_a^{-1} q}\right)}. \tag{A.17}$$

In order to simplify this expression, it is helpful to write $\sqrt{q}^3$ as

$$\sqrt{q}^3 = \sqrt{q}^3 - q\sqrt{T_a q} + q\sqrt{T_a q}$$
$$= q\sqrt{T_a q} - \epsilon\, q\, \Delta_a^{\mathrm{f}} \sqrt{q}$$

$$\begin{aligned}
&= q\sqrt{T_a q} - \sqrt{q}\sqrt{T_a^{-1} q}\sqrt{T_a q} + \sqrt{q}\sqrt{T_a^{-1} q}\sqrt{T_a q} - \epsilon\, q \Delta_a^{\mathrm{f}}\sqrt{q} \\
&= \sqrt{q}\sqrt{T_a^{-1} q}\sqrt{T_a q} + \epsilon\sqrt{q}\sqrt{T_a q}\Delta_a^{\mathrm{b}}\sqrt{q} - \epsilon\, q\Delta_a^{\mathrm{f}}\sqrt{q} \\
&= \sqrt{q}\sqrt{T_a^{-1} q}\sqrt{T_a q} + \epsilon\sqrt{q}\sqrt{T_a q}\Delta_a^{\mathrm{b}}\sqrt{q} - \epsilon\, q\Delta_a^{\mathrm{f}}\sqrt{q} \\
&\quad - \frac{1}{2}\left(\sqrt{T_a q} - \sqrt{T_a^{-1} q}\right)\sqrt{T_a^{-1} q}\sqrt{T_a q} + \frac{1}{2}\left(\sqrt{T_a q} - \sqrt{T_a^{-1} q}\right)\sqrt{T_a^{-1} q}\sqrt{T_a q} \\
&= \frac{1}{2}\left(\sqrt{T_a q} - \sqrt{T_a^{-1} q}\right)\sqrt{T_a^{-1} q}\sqrt{T_a q} + \epsilon\frac{1}{2}\sqrt{T_a^{-1} q}\sqrt{T_a q}\left(-\Delta_a^{\mathrm{f}}\sqrt{q} + \Delta_a^{\mathrm{b}}\sqrt{q}\right) \\
&\quad + \epsilon\sqrt{q}\sqrt{T_a q}\Delta_a^{\mathrm{b}}\sqrt{q} - \epsilon\, q\Delta_a^{\mathrm{f}}\sqrt{q}.
\end{aligned} \tag{A.18}$$

Then, we expand the fraction in equation (A.17) by $\sqrt{q}^3$ and replace $\sqrt{q}^3$ in the numerator by the expression in equation (A.18). This gives

$$\Delta_a \frac{1}{\sqrt{q}} = -\frac{\Delta_a q}{2\sqrt{q}^3} + \epsilon\mu\sqrt{q}_a^{-1}, \tag{A.19}$$

which looks like the continuum result plus a lattice error term given in equation (A.15). $\square$

**Lemma A.5** (Generalized Ricci Identity). *Let $T^{c_1\ldots c_r}{}_{d_1\ldots d_s}$ be an $(r,s)$–lattice tensor. Then the commutator of two covariant lattice differences acting upon $T$ is*

$$\begin{aligned}
[D_a, D_b] T^{c_1\ldots c_r}{}_{d_1\ldots d_s} &= R^{c_1}{}_{fab} T^{fc_2\ldots c_r}{}_{d_1\ldots d_s} + \cdots + R^{c_r}{}_{fab} T^{c_1\ldots c_{r-1}f}{}_{d_1\ldots d_s} \\
&\quad - R^f{}_{d_1 ab} T^{c_1\ldots c_r}{}_{fd_2\ldots d_s} - \cdots - R^f{}_{d_s ab} T^{c_1\ldots c_r}{}_{d_1\ldots d_{s-1}f} \\
&\quad + \epsilon\mu RI(T^{c_1\ldots c_r}{}_{d_1\ldots d_s})_{ab},
\end{aligned} \tag{A.20}$$

*where*

$$\begin{aligned}
\mu RI(T^{c_1\ldots c_r}{}_{d_1\ldots d_s})_{ab} &= \left(\mathcal{O}_a(\Gamma^{c_1}{}_{fb}, T^{fc_2\ldots c_r}{}_{d_1\ldots d_s}) - \mathcal{O}_b(\Gamma^{c_1}{}_{fa}, T^{fc_2\ldots c_r}{}_{d_1\ldots d_s})\right) + \ldots \\
&\quad + \left(\mathcal{O}_a(\Gamma^{c_r}{}_{fb}, T^{c_1\ldots c_{r-1}f}{}_{d_1\ldots d_s}) - \mathcal{O}_b(\Gamma^{c_r}{}_{fa}, T^{c_1\ldots c_{r-1}f}{}_{d_1\ldots d_s})\right) \\
&\quad - \left(\mathcal{O}_a(\Gamma^f{}_{d_1 b}, T^{c_1\ldots c_r}{}_{fd_2\ldots d_s}) - \mathcal{O}_b(\Gamma^f{}_{d_1 a}, T^{c_1\ldots c_r}{}_{fd_2\ldots d_s})\right) - \ldots \\
&\quad - \left(\mathcal{O}_a(\Gamma^f{}_{d_s b}, T^{c_1\ldots c_r}{}_{d_1\ldots d_{s-1}f}) - \mathcal{O}_b(\Gamma^f{}_{d_s a}, T^{c_1\ldots c_r}{}_{d_1\ldots d_{s-1}f})\right).
\end{aligned} \tag{A.21}$$

*Proof.* We show lemma A.5 for a vector and a covector, and the case of a generic tensor follows from representing the tensor as a tensor product of vectors and covectors together with the product rule of finite differences. For a lattice vector $\mathcal{V}^c$, the commutator of covariant differences on the lattice reads

$$\begin{aligned}
[D_a, D_b]\mathcal{V}^c &= D_a(\Delta_b \mathcal{V}^c + \Gamma^c{}_{db}\mathcal{V}^d) - D_b(\Delta_a \mathcal{V}^c + \Gamma^c{}_{da}\mathcal{V}^d) \\
&= \Delta_a(\Delta_b \mathcal{V}^c + \Gamma^c{}_{db}\mathcal{V}^d) - \Gamma^e{}_{ba}(\Delta_e \mathcal{V}^c + \Gamma^c{}_{de}\mathcal{V}^d) + \Gamma^c{}_{ea}(\Delta_b \mathcal{V}^e + \Gamma^e{}_{db}\mathcal{V}^d) \\
&\quad - \Delta_b(\Delta_a \mathcal{V}^c + \Gamma^c{}_{da}\mathcal{V}^d) + \Gamma^e{}_{ab}(\Delta_e \mathcal{V}^c + \Gamma^c{}_{de}\mathcal{V}^d) - \Gamma^c{}_{eb}(\Delta_a \mathcal{V}^e + \Gamma^e{}_{da}\mathcal{V}^d).
\end{aligned} \tag{A.22}$$

Using that finite differences commute and that the lattice Christoffel symbol is symmetric in its lower indices, we get

$$\begin{aligned}[D_a, D_b]\mathcal{V}^c &= \Delta_a(\Gamma^c{}_{db}\mathcal{V}^d) + \Gamma^c{}_{ea}(\Delta_b\mathcal{V}^e + \Gamma^e{}_{db}\mathcal{V}^d) - \Delta_b(\Gamma^c{}_{da}\mathcal{V}^d) - \Gamma^c{}_{eb}(\Delta_a\mathcal{V}^e + \Gamma^e{}_{da}\mathcal{V}^d) \\ &= \left(\Delta_a\Gamma^c{}_{db} - \Delta_b\Gamma^c{}_{da} + \Gamma^c{}_{ea}\Gamma^e{}_{db} - \Gamma^c{}_{eb}\Gamma^e{}_{da}\right)\mathcal{V}^d + \epsilon\left(\mathcal{O}_a(\Gamma^c{}_{db}, \mathcal{V}^d) - \mathcal{O}_b(\Gamma^c{}_{da}, \mathcal{V}^d)\right) \\ &= R^c{}_{dab}\mathcal{V}^d + \epsilon\left(\mathcal{O}_a(\Gamma^c{}_{db}, \mathcal{V}^d) - \mathcal{O}_b(\Gamma^c{}_{da}, \mathcal{V}^d)\right).\end{aligned} \tag{A.23}$$

With this let us define

$$\mu RI(\mathcal{V}^c)_{ab} := \mathcal{O}_a(\Gamma^c{}_{db}, \mathcal{V}^d) - \mathcal{O}_b(\Gamma^c{}_{da}, \mathcal{V}^d) \tag{A.24}$$

which proves the statement for the vector case. For a covector $\mathcal{V}_c$, the proof is very similar. In a first step, we write the commutator of the covariant differences as

$$[D_a, D_b]\mathcal{V}_c = D_a(\Delta_b\mathcal{V}_c - \Gamma^d{}_{cb}\mathcal{V}_d) - D_b(\Delta_a\mathcal{V}_c + \Gamma^d{}_{ca}\mathcal{V}_d). \tag{A.25}$$

$$\tag{A.26}$$

Then, the same proceeding as before yields

$$[D_a, D_b]\mathcal{V}_c = -R^d{}_{cab}\mathcal{V}_d + \epsilon\,\mu RI(\mathcal{V}_c)_{ab} \tag{A.27}$$

where

$$\mu RI(\mathcal{V}_c)_{ab} = -\mathcal{O}_a(\Gamma^d{}_{cb}, \mathcal{V}_d) + \mathcal{O}_b(\Gamma^d{}_{ca}, \mathcal{V}_d). \tag{A.28}$$

□

**Lemma A.6** (Riemann First Skew Symmetry). *The lattice Riemann tensor with lower indices $R_{abcd} := q_{af}R^f{}_{bcd}$ is skew symmetric up to errors of order $\epsilon$ with respect to permuting its anterior indices. Specifically, we have*

$$R_{abcd} = -R_{bacd} + \epsilon\,\mu R_{a\leftrightarrow bcd}, \tag{A.29}$$

*where*

$$\mu R_{a\leftrightarrow bcd} := \mathcal{O}_d(q_{af}, \Gamma^f{}_{cb}) - \mathcal{O}_c(q_{af}, \Gamma^f{}_{db}) + \mathcal{O}_d(q_{bf}, \Gamma^f{}_{ca}) - \mathcal{O}_c(q_{bf}, \Gamma^f{}_{da}). \tag{A.30}$$

*Proof.* $R_{abcd}$ is defined as

$$R_{abcd} = q_{af}\left(\Delta_c\Gamma^f{}_{db} - \Delta_d\Gamma^f{}_{cb} + \Gamma^f{}_{cg}\Gamma^g{}_{db} - \Gamma^f{}_{dg}\Gamma^g{}_{cb}\right). \tag{A.31}$$

Using the product rule to shift the finite differences yields

$$\begin{aligned}R_{abcd} = \Delta_c\Gamma_{adb} &- (\Delta_c q_{af})\Gamma^f{}_{db} - \Delta_d\Gamma_{acb} + (\Delta_d q_{af})\Gamma^f{}_{cb} + \Gamma_{acg}\Gamma^g{}_{db} - \Gamma_{adb}\Gamma^g{}_{cb} \\ &- \epsilon\,\mathcal{O}_c\left(q_{af}, \Gamma^f{}_{db}\right) + \epsilon\,\mathcal{O}_d\left(q_{af}, \Gamma^f{}_{cb}\right),\end{aligned} \tag{A.32}$$

where the Christoffel symbols of the first kind are

$$\Gamma_{adb} := q_{af}\Gamma^f{}_{db} = \Delta_d q_{ab} + \Delta_b q_{da} - \Delta_a q_{db}. \tag{A.33}$$

We insert this expression as well as the explicit formula for the Christoffel symbols of the second kind $\Gamma^f{}_{db}$ into equation (A.32). The result consists of several contributions of the form $\Delta_c \Delta_d q_{ab}$ and $(\Delta_c q_{af})q^{fg}(\Delta_d q_{gb})$ (with varying indices). The same assessment of $R_{bacd}$, adding it to $R_{abcd}$ and comparing terms yields in a straightforward but tedious calculation

$$R_{abcd} + R_{bacd} = \epsilon\left(\mathcal{O}_d(q_{af},\Gamma^f{}_{cb}) - \mathcal{O}_c(q_{af},\Gamma^f{}_{db}) + \mathcal{O}_d(q_{bf},\Gamma^f{}_{ca}) - \mathcal{O}_c(q_{bf},\Gamma^f{}_{da})\right) \tag{A.34}$$

which completes the proof. $\square$

**Lemma A.7** (Riemann Second Skew Symmetry). *The lattice Riemann tensor with lower indices $R_{abcd} := q_{af}R^f{}_{bcd}$ is skew symmetric with respect to permuting its posterior indices*

$$R_{abcd} = -R_{abdc}. \tag{A.35}$$

*Proof.* $R_{abcd}$ is by definition

$$R_{abcd} = q_{af}\left(\Delta_c\Gamma^f{}_{bd} - \Delta_d\Gamma^f{}_{bc} + \Gamma^f{}_{cg}\Gamma^g{}_{bd} - \Gamma^f{}_{dg}\Gamma^g{}_{bc}\right). \tag{A.36}$$

We immediately see that $R_{abcd}$ is antisymmetric upon exchanging the indices $c$ and $d$. Hence, we have $R_{abcd} - R_{abdc} = 0$. $\square$

**Lemma A.8** (Riemann Interchange Symmetry). *The lattice Riemann tensor with lower indices $R_{abcd} := q_{af}R^f{}_{bcd}$ is symmetric upon interchanging the two pairs of anterior and posterior indices up to errors of order $\epsilon$ as given by*

$$R_{abcd} = R_{cdab} + \epsilon\mu R_{(ab)\leftrightarrow(cd)}, \tag{A.37}$$

*where*

$$\mu R_{(ab)\leftrightarrow(cd)} = \mathcal{O}_d(q_{af},\Gamma^f{}_{cb}) - \mathcal{O}_c(q_{af},\Gamma^f{}_{db}) + \mathcal{O}_a(q_{cf},\Gamma^f{}_{bd}) - \mathcal{O}_b(q_{cf},\Gamma^f{}_{ad}). \tag{A.38}$$

*Proof.* The proof requires the very same steps as the proof of lemma A.6. Due to different indices, the lattice terms are also different and given in equation (A.38). $\square$

**Lemma A.9** (Jacobi Identity). *The Jacobi identity holds on the lattice for any lattice tensor $T$,*

$$([D_a,[D_b,D_c]] + [D_b,[D_c,D_a]] + [D_c,[D_a,D_b]])T = 0. \tag{A.39}$$

*Proof.* The proof is trivial – it simply requires to evaluate the commutator brackets. $\square$

**Lemma A.10** (First Bianchi Identity). *The first (algebraic) Bianchi identity holds on the lattice*

$$R_{abcd} + R_{acdb} + R_{adbc} = 0. \tag{A.40}$$

*Proof.* The proof is trivial. Simply write $R_{abcd}$ as

$$R_{abcd} = q_{af}\left(\Delta_c \Gamma^f{}_{bd} - \Delta_d \Gamma^f{}_{cb} + \Gamma^f{}_{gc}\Gamma^g{}_{bd} - \Gamma^f{}_{gd}\Gamma^g{}_{bd}\right) \tag{A.41}$$

and similarly for $R_{acdb}$ and $R_{adbc}$. Then, we add these contributions and compare terms by exploiting the symmetry of $\Gamma$ upon interchanging its lower indices. All contributions cancel. $\square$

**Lemma A.11** (Second Bianchi Identity). *The second Bianchi identity holds on the lattice up to lattice contributions proportional to $\epsilon$. Specifically, we have*

$$D_a R^d{}_{fbc} + D_b R^d{}_{fca} + D_c R^d{}_{fab} = \epsilon \mu BI^d{}_{fabc}, \tag{A.42}$$

*where*

$$\begin{aligned}\mu BI^d{}_{fabc} &= \mathcal{O}_a(\Gamma^d{}_{cg}, \Gamma^g{}_{fb}) - \mathcal{O}_a(\Gamma^d{}_{bg}, \Gamma^g{}_{fc}) + \mathcal{O}_b(\Gamma^d{}_{ag}, \Gamma^g{}_{fc}) - \mathcal{O}_b(\Gamma^d{}_{cg}, \Gamma^g{}_{fa}) \\ &\quad + \mathcal{O}_c(\Gamma^d{}_{bg}, \Gamma^g{}_{fb}) - \mathcal{O}_c(\Gamma^d{}_{ag}, \Gamma^g{}_{fb}).\end{aligned} \tag{A.43}$$

*Proof.* We apply the Jacobi identity lemma A.9 to a lattice vector $\mathcal{V}^d$, and exploit the generalized Ricci identity. The first two contributions to the Jacobi identity are respectively given by

$$\begin{aligned}D_a([D_b, D_c]\mathcal{V}^d) &= D_a\left(R^d{}_{fbc}\mathcal{V}^f + \epsilon \mu RI(\mathcal{V}^d)_{bc}\right) \\ &= (D_a R^d{}_{fbc})\mathcal{V}^d + R^d{}_{fbc} D_a \mathcal{V}^f + \epsilon \left(\mathcal{O}_a(R^d{}_{fbc}, \mathcal{V}^f) + D_a(\mu RI(\mathcal{V}^d)_{bc})\right),\end{aligned} \tag{A.44}$$

$$-[D_b, D_c](D_a \mathcal{V}^d) = -\left(-R^f{}_{abc} D_f \mathcal{V}^d + R^d{}_{fbc} D_a \mathcal{V}^f + \epsilon \mu RI(D_a \mathcal{V}^d)_{bc}\right). \tag{A.45}$$

Upon comparing the two contributions, we see that the remaining terms are

$$(D_a R^d{}_{fbc})\mathcal{V}^f + R^f{}_{abc} D_f \mathcal{V}^d + \epsilon \left(\mathcal{O}_a(R^d{}_{fbc}, \mathcal{V}^f) + D_a(\mu RI(\mathcal{V}^d)_{bc}) - \mu RI(D_a \mathcal{V}^d)_{bc}\right). \tag{A.46}$$

If we evaluate the remaining four contributions to the Jacobi identity, we get similar results with different indices which sum up to

$$0 = \left(D_a R^d{}_{fbc} + D_b R^d{}_{fca} + D_c R^d{}_{fab}\right)\mathcal{V}^f + \left(R^f{}_{abc} + R^f{}_{bca} + R^f{}_{cab}\right) D_f \mathcal{V}^d - \epsilon \mu BI^d{}_{fabc}\mathcal{V}^f \tag{A.47}$$

The second bracket is a version of the first Bianchi identity which holds exactly on the lattice, and hence the contribution cancels. Let us examine the additional lattice contribution $\mu BI^d{}_{fabc}\mathcal{V}^f$. Therefore, we note that the additional lattice term of the product rule defined in equation (3.15) can also take the form

$$\epsilon \mathcal{O}_a(f, g) = \Delta_a(fg) - (\Delta_a f)g - f\Delta_a g. \tag{A.48}$$

We first examine $\epsilon D_a(\mu RI(\mathcal{V}^d)_{bc})$ and express it in the following way:

$$\epsilon D_a(\mu RI(\mathcal{V}^d)_{bc}) \tag{A.49}$$

$$= D_a\big(\epsilon\, \mathcal{O}_b\big(\Gamma^d{}_{cg}, \mathcal{V}^g\big) - \epsilon\, \mathcal{O}_c\big(\Gamma^d{}_{bg}, \mathcal{V}^g\big)\big)$$
$$= D_a\big(\Delta_b(\Gamma^d{}_{cg}\mathcal{V}^g) - \Delta_b(\Gamma^d{}_{cg})\mathcal{V}^g - \Gamma^d{}_{cg}\Delta_b(\mathcal{V}^g) - \Delta_c(\Gamma^d{}_{bg}\mathcal{V}^g) + \Delta_c(\Gamma^d{}_{bg})\mathcal{V}^g + \Gamma^d{}_{bg}\Delta_c(\mathcal{V}^g)\big).$$

The next step consists in evaluating the covariant difference $D_a$ in this equation for the given $(1,2)$–tensor, and which yields a large number of contributions. Many of these cancel with the other two terms of the same form $\epsilon D_b(\mu RI(\mathcal{V}^d)_{ca})$ and $\epsilon D_c(\mu RI(\mathcal{V}^d)_{ab})$. The second kind of lattice terms reads

$$\epsilon\mu RI(D_a\mathcal{V}^d)_{bc} = \epsilon\Big(-\mathcal{O}_b\big(\Gamma^f{}_{ca}, D_f\mathcal{V}^d\big) + \mathcal{O}_c\big(\Gamma^f{}_{ba}, D_f\mathcal{V}^d\big) + \mathcal{O}_b\big(\Gamma^d{}_{cf}, D_a\mathcal{V}^f\big) - \mathcal{O}_c\big(\Gamma^d{}_{bf}, D_a\mathcal{V}^f\big)\Big). \tag{A.50}$$

We express again the individual contributions as in equation (A.48), and get a large number of terms. Many of these cancel with the two other terms of the same form, specifically with $\epsilon\mu RI(D_b\mathcal{V}^d)_{ca}$ and $\epsilon\mu RI(D_c\mathcal{V}^d)_{ab}$. Others cancel with contributions from the first kind of terms from above. For both kinds of terms, we are eventually left with

$$\big(-\Delta_a(R^d{}_{fbc}\mathcal{V}^f) + R^d{}_{fbc}\Delta_a\mathcal{V}^f + \mathcal{V}^f\big(\Gamma^g{}_{fc}\Delta_a\Gamma^d{}_{bg} + \Gamma^d{}_{bg}\Delta_a\Gamma^g{}_{fc} - \Gamma^g{}_{fb}\Delta_a\Gamma^d{}_{cg} - \Gamma^d{}_{cg}\Delta_a\Gamma^g{}_{fb}\big)\big)$$
$$+ (\ -- (a\,b\,c)\ --\ ) + (\ -- (a\,c\,b)\ --\ ), \tag{A.51}$$

where the second line means that the first line appears again but with indices $a,b,c$ cyclically permuted. The $\Gamma$–contributions look like the application of the product rule, and using the latter one can easily establish that equation (A.51) is equivalent to

$$\big(-\Delta_a(R^d{}_{fbc}\mathcal{V}^f) + R^d{}_{fbc}\Delta_a\mathcal{V}^f + \mathcal{V}^f\Delta_aR^d{}_{fbc} + \epsilon\big(\mathcal{O}_a\big(\Gamma^d{}_{cg}, \Gamma^g{}_{fb}\big) - \mathcal{O}_a\big(\Gamma^d{}_{bg}, \Gamma^g{}_{fc}\big)\big)\mathcal{V}^f\big)$$
$$+ (\ -- (a\,b\,c)\ --\ ) + (\ -- (a\,c\,b)\ --\ ). \tag{A.52}$$

The first three contributions simplify to $-\epsilon\,\mathcal{O}_a(R^d{}_{fbc}, \mathcal{V}^f)$, and similarly for the terms with permuted indices. Upon comparing with the total additional lattice term in equation (A.46), we see that this contribution just cancels with the first lattice term there. Hence, we are left with the following expression for the total lattice term

$$\epsilon\big(\mathcal{O}_a\big(\Gamma^d{}_{cg}, \Gamma^g{}_{fb}\big) - \mathcal{O}_a\big(\Gamma^d{}_{bg}, \Gamma^g{}_{fc}\big)\big)\mathcal{V}^f + (\ -- (a\,b\,c)\ --\ ) + (\ -- (a\,c\,b)\ --\ ). \tag{A.53}$$

and which completes the proof. □

**Lemma A.12** (Contracted Bianchi Identity). *The contraction of the second Bianchi identity on the lattice reads*

$$D_b R^b{}_a = \frac{1}{2}D_a R + \epsilon\mu CBI_a, \tag{A.54}$$

*where*

$$\mu CBI_a := \frac{1}{2}\Big(q^{fb}\mu BI^c{}_{fabc} + D_c(q^{fb}q^{cd}\mu R_{f\leftrightarrow dba}) - \mathcal{O}_a(q^{fb}, R_{fb}) - \mathcal{O}_b(q^{fb}, R_{fa}) - \mathcal{O}_c(q^{fb}, R^c{}_{fab})$$
$$-\mu Dq_a{}^{fb}R_{fb} + \mu Dq_b{}^{fb}R_{fa} + \mu Dq_c{}^{fb}R^c{}_{fab}\Big). \tag{A.55}$$

*Proof.* Consider the following contraction of the second Bianchi identity (lemma A.11):

$$q^{fb}\delta^c_d\left(D_a R^d{}_{fbc} + D_b R^d{}_{fca} + D_c R^d{}_{fab}\right) = \epsilon\,\mu BI^d{}_{fabc}q^{fb}\delta^c_d. \tag{A.56}$$

The first contribution gives upon using the posterior skew symmetry of the Riemann tensor and the (inverse) product rule

$$\begin{aligned}q^{fb}D_a R^c{}_{fbc} &= -q^{fb}D_a R_{fb} \\ &= -D_a R + \epsilon\left(\mu Dq_a{}^{fb}R_{fb} + \mathcal{O}_a(q^{fb}, R_{fb})\right).\end{aligned} \tag{A.57}$$

The second contribution yields

$$q^{fb}D_b R_{fa} = D_b R^b{}_a - \epsilon\left(\mu Dq_b{}^{fb}R_{fa} + \mathcal{O}_b(q^{fb}, R_{fa})\right), \tag{A.58}$$

and the third contribution gives using the skew symmetry of the Riemann tensor and the product rule

$$\begin{aligned}q^{fb}D_c R^c{}_{fab} &= D_c(q^{fb}q^{cd}R_{dfab}) - \epsilon\left(\mu Dq_c{}^{fb}R^c{}_{fab} + \mathcal{O}_c(q^{fb}, R^c{}_{fab})\right) \\ &= D_c(q^{fb}q^{cd}(R_{fdba} - \epsilon\mu R_{d\leftrightarrow fba})) - \epsilon\left(\mu Dq_c{}^{fb}R^c{}_{fab} + \mathcal{O}_c(q^{fb}, R^c{}_{fab})\right) \\ &= D_c R^c{}_a - \epsilon\left(D_c(\mu R_{d\leftrightarrow fba}q^{fb}q^{cd}) + \mu Dq_c{}^{fb}R^c{}_{fab} + \mathcal{O}_c(q^{fb}, R^c{}_{fab})\right).\end{aligned} \tag{A.59}$$

Taking all three contributions and the lattice term from the Bianchi identity, the result is

$$0 = -D_a R + 2 D_b R^b{}_a - 2\epsilon\mu CBI_a \tag{A.60}$$

where $\mu CBI_a$ is given in equation (A.55). This completes the proof. $\square$

## B  Lattice Poisson Bracket

We recall that in the continuum classical field theory, the Poisson bracket of two functionals $C : \mathcal{G} \to \mathbb{C}$ and $H : \mathcal{G} \to \mathbb{C}$, where $\mathcal{G}$ denotes an appropriate space of functions on the underlying $n$–dimensional space manifold $\sigma$, is given for two functions $f, g \in \mathcal{G}$ by

$$\{C(f), H(g)\} = \int_\sigma d^n x \left(\frac{\delta C(f)}{\delta q_{ij}(x)}\frac{\delta H(g)}{\delta p^{ij}(x)} - \frac{\delta C(f)}{\delta p^{ij}(x)}\frac{\delta H(g)}{\delta q_{ij}(x)}\right), \tag{B.1}$$

where $\delta C(f)/\delta q_{ij}(x)$ denotes the functional derivative of $C(f)$ with respect to $q_{ij}(x)$, and similarly for the other contributions. More precisely, the functional derivative $\delta C/\delta q$ of a functional $C$ with respect to a function $q$ is defined through

$$\int_\sigma d^n x\, \frac{\delta C}{\delta q}(x)\,\delta q(x) := \left(\frac{d}{ds}C(q + s\,\delta q)\right)_{s=0}, \tag{B.2}$$

where $\delta q(x) \in \mathcal{G}$. Let us examine this expression for the case that the canonical fields are piecewise constant as described in the main part of the paper. In particular, we assume that

$$\delta q(x) = \sum_{X_1...X_n} \delta q^{X_1...X_n} \chi_{X_1...X_n}(x), \quad q(x) = \sum_{X_1...X_n} q^{X_1...X_n} \chi_{X_1...X_n}(x), \tag{B.3}$$

where the sum runs from 0 to $N-1$ for each lattice direction $X_i$, $i \in 1, \ldots, n$, and where $\delta q, q \in \mathcal{F}(\Lambda, \mathbb{R})$ are lattice fields. Thus, the left–hand side of equation (B.2) reduces to

$$\int_\sigma d^n x \, \frac{\delta C}{\delta q}(x) \, \delta q(x) = \sum_{X_1...X_n} \delta q^{X_1...X_n} \int_\sigma d^n x \, \frac{\delta C}{\delta q}(x) \, \chi_{X_1...X_n}(x) \tag{B.4}$$

$$= \sum_{X_1...X_n} \delta q^{X_1...X_n} \int_{V_{X_1...X_n}} d^n x \, \frac{\delta C}{\delta q}(x), \tag{B.5}$$

where $V_{X_1...X_n}$ denotes the volume of the hypercube associated with the lattice point $X_1 \ldots X_n$. Since the functional derivative of e.g., the constraints, is a function of the canonical fields, it is reasonable to assume that it is piecewise constant as well. Consequently, we define

$$\frac{\delta C}{\delta q}(x) := \sum_{X_1...X_n} \left(\frac{\delta C}{\delta q}\right)^{X_1...X_n} \chi_{X_1...X_n}(x). \tag{B.6}$$

This yields for the left–hand side of equation (B.2):

$$\int_\sigma d^n x \, \frac{\delta C}{\delta q}(x) \, \delta q(x) = \epsilon^n \sum_{X_1...X_n} \left(\frac{\delta C}{\delta q}\right)^{X_1...X_n} \delta q^{X_1...X_n}. \tag{B.7}$$

The right–hand side of equation (B.2) by means of equation (B.3) and the chain rule evaluates to

$$\left(\frac{d}{ds} C(q + s \, \delta q)\right)_{s=0} = \sum_{X_1...X_n} \frac{\partial C}{\partial q^{X_1...X_n}} \delta q^{X_1...X_n}. \tag{B.8}$$

We compare the last two equations term by term and obtain that the functional derivative is given by

$$\frac{\delta C}{\delta q}(x) = \frac{1}{\epsilon^n} \sum_{X_1...X_n} \frac{\partial C}{\partial q^{X_1...X_n}} \chi_{X_1...X_n}(x). \tag{B.9}$$

We replace $q$ bei $q_{ij}$ and $p^{ij}$ respectively, and insert this into the original formula for the Poisson bracket, equation (B.1). We get after evaluating the integral the following expression for the Poisson bracket for piecewise constant fields:

$$\{C(f), H(g)\} = \frac{1}{\epsilon^n} \sum_{X_1...X_n} \left( \frac{\partial C(f)}{\partial q_{ij}^{X_1...X_n}} \frac{\partial H(g)}{\partial p^{ij}_{X_1...X_n}} - \frac{\partial C(f)}{\partial p^{X_1...X_n}_{ij}} \frac{\partial H(g)}{\partial q_{ij}^{X_1...X_n}} \right) \tag{B.10}$$

where $C(f), H(g)$ are to be understood as functions of the lattice tensor degrees of freedom $q_{ab}^{Y_1...Y_n}$ and $p^{ab}_{Y_1...Y_n}$ at each $Y_1 \ldots Y_n$.

# C Constraint Algebra on the Lattice

In this appendix, we provide more details on the derivation of the lattice hypersurface deformation algebra as presented in section 5.

## C.1 $\{C(\mathcal{V}), C(\mathcal{W})\}$

For computing the Poisson bracket of two smeared diffeomorphism constraints, we start by deriving the derivative of $C(\mathcal{V})$ with respect to $q_{ij}$ (given in equation (5.9)), where $\mathcal{V}$ is an arbitrary lattice smearing vector. For this purpose, we use that the diffeomorphism constraint on the lattice takes the form

$$C(\mathcal{V}) = \epsilon^n \sum \left(2(\Delta_b \mathcal{V}^a) q_{ac} p^{cb} - \Delta_a(\mathcal{V}^a p^{bc}) q_{bc}\right) \tag{C.1}$$

(see also equation (5.8)), and that we have for the derivative of $q_{ab}$ with respect to $q_{ij}$

$$\frac{\partial q_{ab}^{Y_1 \ldots Y_n}}{\partial q_{ij}^{X_1 \ldots X_n}} = \delta_a^{(i} \delta_b^{j)} \delta_{X_1}^{Y_1} \ldots \delta_{X_n}^{Y_n} \tag{C.2}$$

where all $\delta$'s are Kronecker deltas and the round brackets denote symmetrization with respect to the given indices. Now let us reintroduce lattice indices and relabel the tensor indices to get

$$\begin{aligned}
\frac{\partial C(\mathcal{V})}{\partial q_{ij}^{X_1 \ldots X_n}} &= \frac{\partial}{\partial q_{ij}^{X_1 \ldots X_n}} \epsilon^n \sum_{Y_1 \ldots Y_n = 0}^{N-1} \left(2(\Delta_c \mathcal{V}^a) p^{bc} - \Delta_c(\mathcal{V}^c p^{ab})\right)^{Y_1 \ldots Y_n} q_{ab}^{Y_1 \ldots Y_n} \\
&= \epsilon^n \sum_{Y_1 \ldots Y_n = 0}^{N-1} \left(2(\Delta_c \mathcal{V}^a) p^{bc} - \Delta_c(\mathcal{V}^c p^{ab})\right)^{Y_1 \ldots Y_n} \delta_a^{(i} \delta_b^{j)} \delta_{X_1}^{Y_1} \ldots \delta_{X_n}^{Y_n}
\end{aligned} \tag{C.3}$$

Evaluating the Kronecker deltas yields the following result:

$$\frac{\partial C(\mathcal{V})}{\partial q_{ij}} = \epsilon^n \left((\Delta_c \mathcal{V}^i) p^{jc} + (\Delta_c \mathcal{V}^j) p^{ic} - \Delta_c(\mathcal{V}^c p^{ij})\right). \tag{C.4}$$

The very same strategy applies for deriving $\partial C(\mathcal{V})/\partial p^{ij}$. Therefore, we write the diffeomorphism constraint as

$$C(\mathcal{V}) = \epsilon^n \sum \left(2(\Delta_b \mathcal{V}^a) q_{ac} p^{cb} + \mathcal{V}^a p^{bc}(\Delta_a q_{bc})\right) \tag{C.5}$$

$$\tag{C.6}$$

and perform similar steps as above. This yields

$$\frac{\partial C(\mathcal{V})}{\partial p^{ij}} = \epsilon^n \left((\Delta_j \mathcal{V}^a) q_{ai} + (\Delta_i \mathcal{V}^a) q_{aj} + \mathcal{V}^a(\Delta_a q_{ij})\right) = (\mathcal{L}_\mathcal{V} q)_{ij}. \tag{C.7}$$

The last identity follows by comparing the result with our definition of the Lie difference in definition 4.14. We use equations (C.4) and (C.7) in order to derive the Poisson bracket of $C(\mathcal{V})$ and $C(\mathcal{W})$. Therefore, let us write the first contribution to the Poisson bracket as follows:

$$\epsilon^{-n} \sum \frac{\partial C(\mathcal{V})}{\partial q_{ij}} \frac{\partial C(\mathcal{W})}{\partial p^{ij}} = \epsilon^n \sum \left( (\Delta_a \mathcal{V}^i) p^{aj} + (\Delta_a \mathcal{V}^j) p^{ai} - \Delta_a (\mathcal{V}^a p^{ij}) \right) (\mathcal{L}_\mathcal{W} q)_{ij}. \tag{C.8}$$

By relabeling indices of the two first contributions and using summation by parts for shifting the finite difference in the last term, it immediately follows that

$$\epsilon^{-n} \sum \frac{\partial C(\mathcal{V})}{\partial q_{ij}} \frac{\partial C(\mathcal{W})}{\partial p^{ij}} = \epsilon^n \sum p^{ij} \Big( (\Delta_j \mathcal{V}^a)(\mathcal{L}_\mathcal{W} q)_{ai} + (\Delta_a \mathcal{V}^j)(\mathcal{L}_\mathcal{W} q)_{aj} + \mathcal{V}^a \Delta_a (\mathcal{L}_\mathcal{W} q)_{ij} \Big)$$
$$= \epsilon^n \sum p^{ij} (\mathcal{L}_\mathcal{V} (\mathcal{L}_\mathcal{W} q))_{ij}, \tag{C.9}$$

where the last identity follows by simply looking at definition 4.14 for the Lie difference. Hence, in total we have

$$\{C(\mathcal{V}), C(\mathcal{W})\} = \epsilon^n \sum p^{ij} \Big( \mathcal{L}_\mathcal{V}(\mathcal{L}_\mathcal{W} q)_{ij} - \mathcal{L}_\mathcal{V}(\mathcal{L}_\mathcal{W} q)_{ij} \Big). \tag{C.10}$$

In the continuum, the bracket would simply reduce to $(\mathcal{L}_{[\mathcal{V},\mathcal{W}]} q)_{ij}$ as one can see using the properties of the Lie derivative. On the lattice however, additional terms prevent this identity. The procedure to derive these lattice contributions is as follows: We use definition 4.14 for the Lie difference, and evaluate the two Lie differences acting on $q$ respectively in the previous equation. Many of the numerous contributions cancel without further ado. Others include difference operators that act on products of functions so we need to apply the product rule and thereby generate additional lattice terms. The first of these reads

$$B(\mathcal{V}, \mathcal{W}) = \epsilon^n \sum p^{ij} \left( \mathcal{V}^a \left( 2 \mathcal{O}_a (\Delta_i \mathcal{W}^b, q_{bj}) + \mathcal{O}_a (\mathcal{W}^b, \Delta_b q_{ij}) \right) \right), \tag{C.11}$$

while the second is simply $-B(\mathcal{W}, \mathcal{V})$. Two other contributions to the Poisson bracket remain, the first of which is given by

$$\epsilon^n \sum 2 p^i{}_b \left( (\Delta_i \mathcal{V}^a)(\Delta_a \mathcal{W}^b) + \mathcal{V}^a (\Delta_i \Delta_a \mathcal{W}^b) - (\Delta_i \mathcal{W}^a)(\Delta_a \mathcal{V}^b) - \mathcal{W}^a (\Delta_i \Delta_a \mathcal{V}^b) \right) \tag{C.12}$$

To proceed, it helps to identify the vector

$$\mathcal{N}^b := \mathcal{V}^a \Delta_a \mathcal{W}^b - \mathcal{W}^a \Delta_a \mathcal{V}^b \tag{C.13}$$

in this expression and to see that $\Delta_i$ acts on its factors. Consequently, we can simply write the expression as

$$\epsilon^n \sum 2 p^i{}_b \Delta_i \mathcal{N}^b + \epsilon^{n+1} \sum 2 p^i{}_b \left( \mathcal{O}_i(\mathcal{W}^a, \Delta_a \mathcal{V}^b) - \mathcal{O}_i(\mathcal{V}^a, \Delta_a \mathcal{W}^b) \right). \tag{C.14}$$

where we employed the inverse product rule. We apply summation by parts to shift the difference $\Delta_i$ from $\mathcal{N}^b$ onto $p^i{}_b$. Finally, we include the other remaining contribution to the Poisson bracket and get

$$\{C(\mathcal{V}), C(\mathcal{W})\} = \epsilon^n \sum \mathcal{N}^b \left( -2 \Delta_i (p^i{}_b) + p^{ij} \Delta_b q_{ij} \right) + \epsilon A_{CC}(\mathcal{V}, \mathcal{W}). \tag{C.15}$$

where the additional lattice term is given by

$$A_{CC}(\mathcal{V}, \mathcal{W}) = \epsilon^n \sum 2 p^i{}_b \big( \mathcal{O}_i(\mathcal{W}^a, \Delta_a \mathcal{V}^b) - \mathcal{O}_i(\mathcal{V}^a, \Delta_a \mathcal{W}^b) \big) + B(\mathcal{V}, \mathcal{W}) - B(\mathcal{W}, \mathcal{V}) \quad \text{(C.16)}$$

Finally, we use that by definition $\mathcal{N}^b = (\mathcal{L}_\mathcal{V} \mathcal{W})^b$ in order to write

$$\{C(\mathcal{V}), C(\mathcal{W})\} = C(\mathcal{L}_\mathcal{V} \mathcal{W}) + \epsilon A_{CC}(\mathcal{V}, \mathcal{W}). \quad \text{(C.17)}$$

## C.2 $\{H(f), H(g)\}$

Let us start by deriving $\partial H_{\text{pot}}(f)/\partial q_{ij}$ in equation (5.24) since the derivation of the former results of section 5.2 should be clear. We start by applying the product rule and get

$$\frac{\partial H_{\text{pot}}(f)}{\partial q_{ij}} = -\epsilon^n \frac{\partial}{\partial q_{ij}} \sum f \sqrt{q} \, q^{ab} R_{ab} = -\epsilon^n \sum f \left( \frac{\partial \sqrt{q}}{\partial q^{ij}} R + \sqrt{q} \frac{\partial q^{ab}}{\partial q^{ij}} R_{ab} + \sqrt{q} \, q^{ab} \frac{\partial R_{ab}}{\partial q^{ij}} \right), \quad \text{(C.18)}$$

where we assumed that $q_{ij}$ carries lattice indices $X_1 \ldots X_n$ while all other lattice functions and the sum carry lattice indices $Y_1 \ldots Y_n$. To proceed, we can use similar formulas for the derivatives as in the continuum case, specifically

$$\frac{\partial \sqrt{q^{Y_1 \ldots Y_n}}}{\partial q^{ij}_{X_1 \ldots X_n}} = \frac{1}{2} \sqrt{q^{Y_1 \ldots Y_n}} q^{ij}_{Y_1 \ldots Y_n} \delta^{Y_1}_{X_1} \ldots \delta^{Y_n}_{X_n}, \quad \text{(C.19)}$$

$$\frac{\partial q^{ab}_{Y_1 \ldots Y_n}}{\partial q^{X_1 \ldots X_n}_{ij}} = -\frac{1}{2} \big(q^{ai} q^{bj} + q^{aj} q^{bi}\big)^{Y_1 \ldots Y_n} \delta^{Y_1}_{X_1} \ldots \delta^{Y_n}_{X_n}. \quad \text{(C.20)}$$

Inserting these expressions into the previous relation gives

$$\frac{\partial H_{\text{pot}}(f)}{\partial q_{ij}} = -\frac{\epsilon^n}{2} f \sqrt{q} \big( R q^{ij} - R^{ij} - R^{ji} \big) - \epsilon^n \sum f \sqrt{q} \, q^{ab} \frac{\partial R_{ab}}{\partial q^{ij}}. \quad \text{(C.21)}$$

On the lattice, $R^{ij}$ is not symmetric but the following identity holds by using lemma 4.9

$$R^{ji} = q^{ja} q^{ib} q^{cd} R_{dacb} = q^{ja} q^{ib} q^{cd} \big( R_{cbda} + \epsilon \mu R_{(da)\leftrightarrow(cb)} \big) \quad \text{(C.22)}$$

$$= R^{ij} + \epsilon \, q^{ja} q^{ib} q^{cd} \mu R_{(da)\leftrightarrow(cb)} \quad \text{(C.23)}$$

$$=: R^{ij} + 2\epsilon \, C^{ij}. \quad \text{(C.24)}$$

Hence, we obtain for equation (C.18) by setting $R^{ij} - (1/2) \, q^{ij} R =: G^{ij}$,

$$\frac{\partial H_{\text{pot}}(f)}{\partial q_{ij}} = \epsilon^n f \sqrt{q} \big( G^{ij} + \epsilon \, C^{ij} \big) - \epsilon^n \sum f \sqrt{q} \, q^{ab} \frac{\partial R_{ab}}{\partial q^{ij}}. \quad \text{(C.25)}$$

Since the first contributions are proportional to $f$ (no derivatives appearing), they will cancel in the final expression for the Poisson bracket, and we can consequently ignore them from now on. The term involving the derivative of the lattice Ricci tensor will not cancel. To see this let us establish the following lemma.

**Lemma C.1** (Derivatives and Differences Commute). *Let $f \in \mathcal{F}(\Lambda, \mathbb{R})$ be a lattice function that depends on the metric tensor $q_{ab}$. Then, it holds true that*

$$\frac{\partial (\Delta_a f(q_{ab}))^{Y_1 \ldots Y_n}}{\partial q_{ij}^{X_1 \ldots X_n}} = \Delta_a \left( \frac{\partial f(q_{ab})^{Y_1 \ldots Y_n}}{\partial q_{ij}^{X_1 \ldots X_n}} \right). \quad (C.26)$$

*Proof.* The proof is straightforward by writing out the definition of the difference operator on the lattice. Specifically, we have

$$\frac{\partial (\Delta_a f(q_{ab}))^{Y_1 \ldots Y_n}}{\partial q_{ij}^{X_1 \ldots X_n}} = \frac{\partial}{\partial q_{ij}^{X_1 \ldots X_n}} \frac{1}{2\epsilon} \left( f(q_{ab})^{Y_1 \ldots Y_a + 1 \bmod N \ldots Y_n} - f(q_{ab})^{Y_1 \ldots Y_a - 1 \bmod N \ldots Y_n} \right)$$

$$= \frac{1}{2\epsilon} \left( \frac{\partial f(q_{ab})^{Y_1 \ldots Y_a + 1 \bmod N \ldots Y_n}}{\partial q_{ij}^{X_1 \ldots X_n}} - \frac{\partial f(q_{ab})^{Y_1 \ldots Y_a - 1 \bmod N \ldots Y_n}}{\partial q_{ij}^{X_1 \ldots X_n}} \right)$$

$$= \Delta_a \left( \frac{\partial f(q_{ab})^{Y_1 \ldots Y_n}}{\partial q_{ij}^{X_1 \ldots X_n}} \right). \quad (C.27)$$

□

This identity motivates us to define a short hand notation for the following proofs. In particular, let us abbreviate the partial derivative with respect to $q_{ij}$ by "$\partial$". With this, let us spell out the following lemmas.

**Lemma C.2** (Derivative of $\Gamma^a{}_{bc}$). *The derivative of the Christoffel symbol with respect to the metric reads*

$$\partial \Gamma^a{}_{bc} = \frac{q^{ad}}{2} (D_c \, \partial q_{bd} + D_b \, \partial q_{cd} - D_d \, \partial q_{bc}). \quad (C.28)$$

*Proof.* We apply the product rule for the derivative of the Christoffel symbol to get

$$\partial \Gamma^a{}_{bc} = \frac{1}{2} (\partial q^{ad} (\Delta_b q_{dc} + \Delta_c q_{bd} - \Delta_d q_{bc}) + q^{ad} (\Delta_b \, \partial q_{dc} + \Delta_c \, \partial q_{bd} - \Delta_d \, \partial q_{bc})) \quad (C.29)$$

We insert the derivative of the inverse metric given in equation (C.20) in this expression. On the other hand, we can employ the definition of the covariant difference operator and get

$$D_c \, \partial q_{bd} = \Delta_c \, \partial q_{bd} - \Gamma^e{}_{bc} \, \partial q_{ed} - \Gamma^e{}_{dc} \, \partial q_{be}. \quad (C.30)$$

Here, we took into account that $\partial q_{bd}$ is a tensor but $D_c \, \partial q_{bd} \neq 0$. We write out the same formula for the other covariant differences appearing in equation (C.28) and add the terms accordingly. Upon comparing the result with equation (C.29), we see that the expressions are identical if we use that

$$\frac{q^{ad}}{2} (-2 \Gamma^e{}_{bc} \, \partial q_{ed}) = -q^{ai} q^{jd} (\Delta_b q_{dc} + \Delta_c q_{bd} - \Delta_d q_{bc}) \, \partial q_{ij}. \quad (C.31)$$

□

**Lemma C.3** (Derivative of $R_{ab}$)**.** *The partial derivative of $R_{ab}$ with respect to the metric tensor is given by*

$$\partial R_{ab} = D_d \, \partial \Gamma^d{}_{ab} - D_a \, \partial \Gamma^d{}_{db}. \tag{C.32}$$

*Proof.* To begin with, let us note that $\partial \Gamma^a{}_{bc}$ is a tensor which can be seen from lemma C.2 since all individual contributions to $\partial \Gamma^a{}_{bc}$ are tensors. Hence we can use the definition of the covariant difference operator to compute $D_c \, \partial \Gamma^a{}_{bc}$ and derive an expression for the combination in equation (C.32). On the other hand, we can employ the definition of $R_{ab}$ as given in equations (5.4) and (5.5) and apply the product rule for the partial derivative with respect to $q_{ij}$ in order to compute $\partial R_{ab}$ directly. Without further manipulations, one can see that the two expressions are identical. □

Given lemmas C.2 and C.3, it is straightforward to establish that

$$\partial R_{ab} = \frac{1}{2}\left(D_c(q^{cd}D_a \, \partial q_{db}) + D_c(q^{cd}D_b \, \partial q_{da}) - D_a(q^{cd}D_b \, \partial q_{cd}) - D_c(q^{cd}D_d \, \partial q_{ab})\right). \tag{C.33}$$

For this, we simply had to insert the expression for $\partial \Gamma^a{}_{bc}$ given in equation (C.28) into equation (C.32). With this result, let us consider the whole expression in which $\partial R_{ab}$ appears and shift the covariant differences appearing in the previous expression to the other lattice fields occuring in the sum. In particular, we derive that

$$-\epsilon^n \sum f\sqrt{q}\, q^{ab} \frac{\partial R_{ab}}{\partial q^{ij}} = \frac{\epsilon^n}{2}\left(D_a\big(D_b\big(f\sqrt{q}q^{ij}\big)q^{ab}\big) + D_a\big(D_b\big(f\sqrt{q}q^{ab}\big)q^{ij}\big)\right)$$
$$- \frac{\epsilon^n}{2}\left(D_a\big(D_b\big(f\sqrt{q}q^{aj}\big)q^{bi}\big) + D_a\big(D_b\big(f\sqrt{q}q^{ai}\big)q^{bj}\big)\right). \tag{C.34}$$

For later convenience, we want the covariant differences to act on $f$ alone. We apply the product rule for finite differences and exemplarily consider one of the contributions above. We write

$$D_a\big(D_b\big(f\sqrt{q}q^{ij}\big)q^{ab}\big) = D_a\big(\big((D_b f)\sqrt{q}q^{ij} + fD_b(\sqrt{q}q^{ij}) + \epsilon\, \mathcal{O}_b(f, \sqrt{q}q^{ij})\big)q^{ab}\big)$$
$$= \sqrt{q}q^{ij}q^{ab}D_a D_b f + \epsilon\, F^{ij}(f, q^{ij}, q^{ab}), \tag{C.35}$$

where we defined for the last step

$$F^{ij}(f, q^{ij}, q^{ab}) = \epsilon^{n-1}\big((D_b f)D_a(\sqrt{q}q^{ij}q^{ab}) + D_a(fq^{ab}D_b(\sqrt{q}q^{ij}))\big)$$
$$+ \epsilon^n \mathcal{O}_a(D_b f, \sqrt{q}q^{ij}q^{ab}) + \epsilon^n D_a(\mathcal{O}_b(f, \sqrt{q}q^{ij})q^{ab}). \tag{C.36}$$

All contributions together yield

$$-\epsilon^n \sum f\sqrt{q}\, q^{ab} \frac{\partial R_{ab}}{\partial q^{ij}} = \epsilon^n \sqrt{q}\,(q^{ij}q^{ab} - q^{ai}q^{bj})D_a(D_b f) + \epsilon\, E^{ij}(f), \tag{C.37}$$

where $E^{ij}(f)$ is defined from $F^{ij}$ as in equation (5.28). The final step for computing the Poisson bracket of two smeared Hamilton constraints consists in multiplying this expression with

$$\frac{\partial H_{\text{kin}}(f)}{\partial p^{ij}} = \epsilon^n f \frac{2}{\sqrt{q}}\Big(p_{ij} - \frac{p}{n-1}q_{ij}\Big). \tag{C.38}$$

It is straightforward to show that this yields in total

$$\epsilon^n \sum 2(fD_aD_bg - gD_aD_bf)p^{ab} + \epsilon^{n+1} \sum \frac{2}{\sqrt{q}}\Big(p_{ij} - \frac{p}{n-1}q_{ij}\Big)\big(gE^{ij}(f) - fE^{ij}(g)\big). \quad \text{(C.39)}$$

Finally, we shift the outer covariant differences in the first contribution and apply the lattice product rule for the covariant difference. A straightforward computation yields the final result in equation (5.36).

## C.3   $\{C(\mathcal{V}), H(f)\}$

The first contribution to the Poisson bracket reads

$$\frac{\partial C(\mathcal{V})}{\partial q_{ij}}\frac{\partial H(f)}{\partial p^{ij}} = \epsilon^n\big((\Delta_d\mathcal{V}^i)p^{jd} + (\Delta_d\mathcal{V}^j)p^{id} - \Delta_d(\mathcal{V}^d p^{ij})\big)\cdot \epsilon^n f\frac{\partial \mathcal{H}_{\text{kin}}}{\partial p^{ij}}, \quad \text{(C.40)}$$

where we used that only the kinetic part of the Hamilton constraint depends on $p^{ij}$. To evaluate this expression, we apply the lattice product rule to the last term and employ the explicit expression

$$\frac{\partial \mathcal{H}_{\text{kin}}}{\partial p^{ij}} = \frac{2}{\sqrt{q}}\Big(p_{ij} - \frac{p}{n-1}q_{ij}\Big). \quad \text{(C.41)}$$

This yields

$$\frac{1}{\epsilon^{2n}}\frac{\partial C(\mathcal{V})}{\partial q_{ij}}\frac{\partial H(f)}{\partial p^{ij}} = \frac{4f(\Delta_a\mathcal{V}^b)}{\sqrt{q}}\Big(p_{bc}p^{ca} - \frac{pp_b{}^a}{n-1}\Big) - 2f(\Delta_a\mathcal{V}^d)\mathcal{H}_{\text{kin}} - f\mathcal{V}^d(\Delta_d p^{ij})\frac{\partial \mathcal{H}_{\text{kin}}}{\partial p^{ij}}$$
$$- \epsilon\,\mathcal{O}_d(\mathcal{V}^d, p^{ij})f\frac{\partial \mathcal{H}_{\text{kin}}}{\partial p^{ij}} \quad \text{(C.42)}$$

Then, consider the second contribution to the Poisson bracket but restricted to the kinetic part of the Hamiltonian. Using the definition of the two factors that enter this term, we get by a straightforward multiplication of these factors

$$-\frac{\partial C(\mathcal{V})}{\partial p^{ij}}f\frac{\partial \mathcal{H}_{\text{kin}}}{\partial q_{ij}} = \epsilon^n f\Bigg((\Delta_a\mathcal{V}^a)\mathcal{H}_{\text{kin}} - \frac{4(\Delta_a\mathcal{V}^b)}{\sqrt{q}}\Big(p_{bc}p^{ca} - \frac{pp_b{}^a}{n-1}\Big) - \mathcal{V}^a(\Delta_a q_{ij})\frac{\partial \mathcal{H}_{\text{kin}}}{\partial q_{ij}}\Bigg). \quad \text{(C.43)}$$

The sum of this expression and the previous equation gives in total

$$\frac{\partial C(\mathcal{V})}{\partial q_{ij}}\frac{\partial H(f)}{\partial p^{ij}} - \frac{\partial C(\mathcal{V})}{\partial p^{ij}}\epsilon^n f\frac{\partial \mathcal{H}_{\text{kin}}}{\partial q_{ij}} = \epsilon^{2n}f\bigg(-(\Delta_d\mathcal{V}^d)\mathcal{H}_{\text{kin}} - \mathcal{V}^d\Big((\Delta_d q_{ij})\frac{\partial \mathcal{H}_{\text{kin}}}{\partial q_{ij}} + (\Delta_d p^{ij})\frac{\partial \mathcal{H}_{\text{kin}}}{\partial p^{ij}}\Big)\bigg)$$
$$- \epsilon^{2n+1}\mathcal{O}_d(\mathcal{V}^d, p^{ij})f\frac{\partial \mathcal{H}_{\text{kin}}}{\partial p^{ij}}. \quad \text{(C.44)}$$

To further simplify this expression, let us use the following lemma.

**Lemma C.4** (Lattice Difference of $\mathcal{H}_{\text{kin}}$). *Let $\mathcal{H}_{\text{kin}}$ be the kinetic part of the lattice Hamiltonian density given by*

$$\mathcal{H}_{kin} = \frac{1}{\sqrt{q}}\left(q_{ac}q_{bd} - \frac{1}{n-1}q_{ab}q_{cd}\right)p^{ab}p^{cd}. \tag{C.45}$$

*Then, the lattice difference of $\mathcal{H}_{\text{kin}}$ reads*

$$\Delta_e \mathcal{H}_{kin} = (\Delta_e p^{ij})\frac{\partial \mathcal{H}_{kin}}{\partial p^{ij}} + (\Delta_e q_{ij})\frac{\partial \mathcal{H}_{kin}}{\partial q_{ij}} - \epsilon\, K_e, \tag{C.46}$$

*where the lattice term $K_e$ is given by*

$$K_e = \left(-\frac{\mu q_e}{2\sqrt{q}^3} + \mu\sqrt{q}_e^{-1}\right)\sqrt{q}\mathcal{H}_{kin} + \frac{p^{bc}p^{df}}{\sqrt{q}}\left(\mathcal{O}_e(q_{bd}, q_{cf}) - \frac{\mathcal{O}_e(q_{bc}, q_{df})}{n-1}\right) + \mathcal{O}_e\left(\frac{1}{\sqrt{q}}, \sqrt{q}\mathcal{H}_{kin}\right)$$
$$+ \frac{1}{\sqrt{q}}\mathcal{O}_e(G_{abcd}, p^{ab}p^{cd}) + \frac{G_{abcd}}{\sqrt{q}}\mathcal{O}_e(p^{ab}, p^{cd}). \tag{C.47}$$

*Proof.* In order to compute $\Delta_e \mathcal{H}_{\text{kin}}$, we apply the product rule for the finite difference to $\mathcal{H}_{\text{kin}}$ as defined in equation (C.45). Applying the finite difference to the first factor of $\mathcal{H}_{\text{kin}}$ gives

$$\Delta_e\left(\frac{1}{\sqrt{q}}\right) = -\frac{\Delta_e q}{2\sqrt{q}^3} + \epsilon\mu\sqrt{q}_e^{-1} = -(\Delta_e q_{bc})\frac{q^{bc}}{2\sqrt{q}} + \epsilon\left(-\frac{\mu q_e}{2\sqrt{q}^3} + \mu\sqrt{q}_e^{-1}\right) \tag{C.48}$$

as we have shown in lemma A.4. When we apply the finite difference to the second factor in equation (C.45) it is useful to contract it with the momenta $p^{ab}p^{cd}$ right away which yields

$$(\Delta_e G_{abcd})p^{ab}p^{cd} = 2(\Delta_e q_{bc})\left(p^b{}_d p^{dc} - \frac{p}{n-1}p^{bc}\right) + \epsilon\, p^{ab}p^{cd}\left(\mathcal{O}_e(q_{ac}, q_{bd}) - \frac{\mathcal{O}_e(q_{ab}, q_{cd})}{n-1}\right). \tag{C.49}$$

Similarly, we get for the finite difference of the momentum part contracted with the supermetric

$$\Delta_e(p^{ab}p^{cd})G_{abcd} = 2\,(\Delta_e p^{ab})\left(p_{ab} - \frac{p}{n-1}q_{ab}\right) + \epsilon\, G_{abcd}\mathcal{O}_e(p^{ab}, p^{cd}). \tag{C.50}$$

Finally, we need to take the lattice terms from applying the product rule to $\mathcal{H}_{\text{kin}}$ earlier into account. As one can easily show, they read

$$\mathcal{O}_e\left(\frac{1}{\sqrt{q}}, \sqrt{q}\mathcal{H}_{kin}\right) + \epsilon\,\frac{1}{\sqrt{q}}\mathcal{O}_e(G_{abcd}, p^{ab}p^{cd}). \tag{C.51}$$

We add all terms and get

$$\Delta_e \mathcal{H}_{kin} = (\Delta_e q_{bc})\frac{\partial \mathcal{H}_{kin}}{\partial q_{bc}} + (\Delta_e p^{bc})\frac{\partial \mathcal{H}_{kin}}{\partial p^{bc}} + \epsilon\, K_e, \tag{C.52}$$

where we employed that

$$\frac{\partial \mathcal{H}_{\text{kin}}}{\partial q_{bc}} = -\frac{q^{bc}}{2}\mathcal{H}_{\text{kin}} + \frac{2}{\sqrt{q}}\left(p^b{}_c p^{dc} - \frac{p}{n-1}p^{bc}\right), \tag{C.53}$$

$$\frac{\partial \mathcal{H}_{\text{kin}}}{\partial p^{bc}} = \frac{2}{\sqrt{q}}\left(p_{bc} - \frac{p}{n-1}q_{bc}\right), \tag{C.54}$$

and where the lattice contribution $K_e$ is as defined in equation (C.47). □

As a consequence, we can simplify the above contributions to the Poisson bracket and get

$$\frac{\partial C(\mathcal{V})}{\partial q_{ij}}\frac{\partial H(f)}{\partial p^{ij}} - \frac{\partial C(\mathcal{V})}{\partial p^{ij}}\epsilon^n f\frac{\partial \mathcal{H}_{\text{kin}}}{\partial q_{ij}} = -\epsilon^{2n}f\big((\Delta_d \mathcal{V}^d)\mathcal{H}_{\text{kin}} + \mathcal{V}^d \Delta_d \mathcal{H}_{\text{kin}}\big)$$
$$+ \epsilon^{2n+1}\left(\mathcal{O}_d(\mathcal{V}^d, p^{ij})f\frac{\partial \mathcal{H}_{\text{kin}}}{\partial p^{ij}} - f\mathcal{V}^d K_d\right) \tag{C.55}$$

This expression suggests to apply the inverse product rule and a summation by parts when the enclosing sum is taken into account. This yields

$$\epsilon^{-n}\sum\left(\frac{\partial C(\mathcal{V})}{\partial q_{ij}}\frac{\partial H(f)}{\partial p^{ij}} - \frac{\partial C(\mathcal{V})}{\partial p^{ij}}\epsilon^n f\frac{\partial \mathcal{H}_{\text{kin}}}{\partial q_{ij}}\right) \tag{C.56}$$
$$= \epsilon^n \sum \mathcal{V}^d(\Delta_d f)\mathcal{H}_{\text{kin}} + \epsilon^{n+1}\sum\left(f\mathcal{O}_d(\mathcal{V}^d, \mathcal{H}_{\text{kin}}) + \mathcal{O}_d(\mathcal{V}^d, p^{ij})f\frac{\partial \mathcal{H}_{\text{kin}}}{\partial p^{ij}} - f\mathcal{V}^d K_d\right)$$

We define the lattice term in this expression by $\epsilon J(f, \mathcal{V})$. As suggested in section 5.3, the final contribution to the Poisson bracket involves the product

$$-\frac{1}{\epsilon^{2n}}\frac{\partial C(\mathcal{V})}{\partial p^{ij}}\frac{\partial H_{\text{pot}}(f)}{\partial q_{ij}} = -\big((D_j\mathcal{V}^a)q_{ai} + (D_i\mathcal{V}^a)q_{aj}\big)\big(f\sqrt{q}G^{ij} + \sqrt{q}\,(q^{ij}q^{ab} - q^{ai}q^{bj})D_a D_b f\big)$$
$$- \epsilon\big((D_j\mathcal{V}^a)q_{ai} + (D_i\mathcal{V}^a)q_{aj}\big)\big(f\sqrt{q}\,C^{ij} + E^{ij}(f)\big). \tag{C.57}$$

In order to simplify notation, we denote the sum over the lattice error term in this expression by

$$M(f, \mathcal{V}) := -\epsilon^n \sum\big((D_j\mathcal{V}^a)q_{ai} + (D_i\mathcal{V}^a)q_{aj}\big)\big(f\sqrt{q}\,C^{ij} + E^{ij}(f)\big). \tag{C.58}$$

Then, let's focus on the terms $\propto G^{ij}$. Using that $G^{ij} = R^{ij} - (1/2)q^{ij}R$, it is straightforward to show that

$$-\big((D_j\mathcal{V}^a)q_{ai} + (D_i\mathcal{V}^a)q_{aj}\big)G^{ij} = -R^j{}_a D_j\mathcal{V}^a - R_a{}^j D_j\mathcal{V}^a + R D_a\mathcal{V}^a. \tag{C.59}$$

In the continuum, the two first contributions are identical due to the symmetry of the Ricci tensor. On the lattice however, this symmetry only holds up to lattice terms which we derive as follows:

$$R_a{}^j = q^{jb}R_{ab} = q^{jb}q^{cd}R_{cadb}$$
$$= q^{jb}q^{cd}\big(R_{dbca} + \epsilon \mu R_{(ca)\leftrightarrow(db)}\big)$$

$$= R^j{}_a + \epsilon\, q^{jb} q^{cd} \mu R_{(ca)\leftrightarrow(db)}. \tag{C.60}$$

For the final result, we would like to shift the covariant differences in equation (C.59) from $\mathcal{V}$ to $f$ when the additional factor $\sqrt{q}f$ is taken into account. Within the enclosing sum, we perform a summation by parts and after a straightforward application of the lattice product rule, we obtain

$$-\epsilon^n \sum f\sqrt{q}\big((D_j\mathcal{V}^a)q_{ai} + (D_i\mathcal{V}^a)q_{aj}\big)G^{ij} = \epsilon^n \sum \big(2\sqrt{q}\mathcal{V}^a R^b{}_a \Delta_b f - \sqrt{q}\mathcal{V}^a R \Delta_a f\big) + \epsilon N(f, \mathcal{V}) \tag{C.61}$$

where the additional lattice contributions $N(f, \mathcal{V})$ are given by

$$N(f, \mathcal{V}) = \epsilon^n \sum \big(f\sqrt{q}q^{jb}q^{cd}\mu R_{(ca)\leftrightarrow(db)} + 2f\mathcal{V}^a \mu D\sqrt{q}_b R^b{}_a - f\mathcal{V}^a R \mu D\sqrt{q}_a\big). \tag{C.62}$$

The last step of our computation concerns the contributions in equation (C.57) proportional to $D_a D_b f$, in particular these are

$$\big((D_j\mathcal{V}^a)q_{ai} + (D_i\mathcal{V}^a)q_{aj}\big)\sqrt{q}\big(q^{ai}q^{bj} - q^{ij}q^{ab}\big)D_a D_b f = 2\sqrt{q}\big(q^{bj}D_j\mathcal{V}^a - q^{ab}D_c\mathcal{V}^c\big)D_a D_b f. \tag{C.63}$$

In order to be able to compare those terms with the other contributions, we have to shift the covariant difference that operates on $\mathcal{V}$ to the lattice function $f$. For this, we perform first a summation by parts and then apply the product rule. This gives

$$\epsilon^n \sum 2\sqrt{q}\big(q^{bj}D_j\mathcal{V}^a - q^{ab}D_c\mathcal{V}^c\big)D_a D_b f = \epsilon^n \sum 2\sqrt{q}\mathcal{V}^c q^{ab}(D_c D_a D_b - D_a D_c D_b)f + \epsilon T(f, \mathcal{V}), \tag{C.64}$$

where the additional lattice term $T(f, \mathcal{V})$ reads

$$T(f, \mathcal{V}) = \epsilon^n \sum 2\mathcal{V}^c\big(\mu D\sqrt{q}_c D^a D_a f - \mu D\sqrt{q}_a q^{ba} D_c D_b f + \sqrt{q}\mu D q_c{}^{ab} D_a D_b f - \sqrt{q}\mu D q_a{}^{ab} D_c D_b f$$
$$+\sqrt{q}\big(\mathcal{O}_c(q^{ab}, D_a D_b f) - \mathcal{O}_a(q^{ab}, D_c D_b f)\big) + \mathcal{O}_c(\sqrt{q}, D^a D_a f) - \mathcal{O}_a(\sqrt{q}, q^{ab} D_c D_b f)\big). \tag{C.65}$$

In a last step, we apply the generalized Ricci identity (lemma 4.6) in order to commute covariant differences in equation (C.64) and apply lemma 4.7 to permute indices of the Riemann tensor. This yields

$$\epsilon^n \sum 2\sqrt{q}\big(q^{bj}D_j\mathcal{V}^a - q^{ab}D_c\mathcal{V}^c\big)D_a D_b f = -\epsilon^n \sum 2\sqrt{q}\mathcal{V}^a R^b{}_a \Delta_b f + \epsilon M(f, \mathcal{V}), \tag{C.66}$$

where $M(f, \mathcal{V})$ is defined by

$$S(f, \mathcal{V}) = \epsilon^n \sum 2\sqrt{q}\mathcal{V}^c q^{ab}\big(q^{df}\mu R_{f\leftrightarrow bac}\Delta_d f - \mu RI(D_b f)_{ac}\big) + T(f, \mathcal{V}). \tag{C.67}$$

We are ready to add all contributions, in particular, equations (C.56), (C.58), (C.61) and (C.66). Two terms cancel and we end up with

$$\{C(\mathcal{V}), H(f)\} = \epsilon^n \sum \big(\mathcal{V}^d(\Delta_d f)\big(\mathcal{H}_{\text{kin}} - \sqrt{q}R\big)\big) + \epsilon A_{CH}(f, \mathcal{V}) \tag{C.68}$$

where we defined

$$A_{CH}(f, \mathcal{V}) := J(f, \mathcal{V}) + M(f, \mathcal{V}) + N(f, \mathcal{V}) + S(f, \mathcal{V}). \tag{C.69}$$

Knowing that $\mathcal{H}_{\text{pot}} = -\sqrt{q}R$, the final result is

$$\begin{aligned} \{C(\mathcal{V}), H(f)\} &= \epsilon^n \sum \mathcal{V}^d(\Delta_d f)\mathcal{H} + \epsilon A_{CH}(f, \mathcal{V}) \\ &= H(\mathcal{L}_\mathcal{V} f) + \epsilon A_{CH}(f, \mathcal{V}) \end{aligned} \tag{C.70}$$

where we employed that $\mathcal{L}_\mathcal{V} f = \mathcal{V}^d \Delta_d f$.